\RequirePackage{fix-cm}
\documentclass[11pt,journal,compsoc,onecolumn]{IEEEtran}      % twocolumn
\usepackage{graphicx}
\newtheorem{Theo}{Theorem}
\usepackage{algorithm}
\usepackage{algorithmic}
\usepackage{subfigure}
\usepackage{comment}
\usepackage{changes}
\usepackage{setspace} %double line spacing
\usepackage{array}

\newcommand{\PreserveBackslash}[1]{\let\temp=\\#1\let\\=\temp}
\newcolumntype{C}[1]{>{\PreserveBackslash\centering}p{#1}}
\newcolumntype{R}[1]{>{\PreserveBackslash\raggedleft}p{#1}}
\newcolumntype{L}[1]{>{\PreserveBackslash\raggedright}p{#1}}

\begin{document}
%\large
\begin{spacing}{1.5} % when submitting to TWC, this note should be commented

\title{Efficient Three-stage Auction Schemes for Cloudlets Deployment in Wireless Access Network\thanks{This article has been accepted by Wireless Networks, contents differ from the final version. This is not the final version. This article is used only for quick dissemination of research findings and only for education purpose.}
}

%\subtitle{Do you have a subtitle?\\ If so, write it here}

%\titlerunning{Short form of title}        % if too long for running head

%\authorrunning{Short form of author list} % if too long for running head

	\author{\IEEEauthorblockN{Gangqiang Zhou\IEEEauthorrefmark{1},
		Jigang Wu\IEEEauthorrefmark{1},  Long Chen\IEEEauthorrefmark{1}, Guiyuan Jiang\IEEEauthorrefmark{2}, Siew-Kei Lam\IEEEauthorrefmark{2} }    \\
	\IEEEauthorblockA{\IEEEauthorrefmark{1}School of Computer Science and Technology\\
		Guangdong University of Technology,	Guangzhou, Guangdong 510006\\
		\IEEEauthorrefmark{2}School of Computer Science and Engineering, Nanyang
		Technological University, Singapore\\
		 Email:gq\_zhou@outlook.com , asjgwucn@outlook.com, lonchen@mail.ustc.edu.cn, gyjiang@ntu.edu.sg, assklam@ntu.edu.sg
		}
}

% make the title area
\maketitle

% The correct dates will be entered by the editor

\maketitle

\begin{abstract}

Cloudlet deployment and resource allocation for mobile users (MUs) have been extensively studied in existing works for computation resource scarcity. However, most of them failed to jointly consider the two techniques together, and the selfishness of cloudlet and access point (AP) are ignored. Inspired by the group-buying mechanism, this paper proposes three-stage auction schemes by combining cloudlet placement and resource assignment, to improve the social welfare subject to the economic properties. We first divide all MUs into some small groups according to the associated APs. Then the MUs in same group can trade with cloudlets in a group-buying way through the APs. Finally, the MUs pay for the cloudlets if they are the winners in the auction scheme. We prove that our auction schemes can work in polynomial time. We also provide the proofs for economic properties in theory. For the purpose of performance comparison, we compare the proposed schemes with HAF, which is a centralized cloudlet placement scheme without auction. Numerical results confirm the correctness and efficiency of the proposed schemes.


% \PACS{PACS code1 \and PACS code2 \and more}
% \subclass{MSC code1 \and MSC code2 \and more}
\end{abstract}

\section{Introduction}
\label{intro}
In recent years, portable devices such as
smartphones and tablet PCs have evolved to
reach a significant performance enhancement.
However, applications running on those
mobile devices also consume many resources, e.g.
computing, storage, et al.
Particularly, multiple applications are often
run on the same devices of mobile users (MUs).

Therefore, the resource-limited mobile devices still
require a lot more resources for better performance,
to tackle real-time and delay-sensitive tasks,
such as Virtual Reality games and Automatic driving.

A cloudlet is formed by a group of internet-well-connected,
resource-rich, and trusted computers
When the centralized cloud is too far away from MUs.
Cloudlet can be utilized by neighboring MUs \cite{Satyanarayanan2009The},
and it also can bring us a good solution for the resource
requirement problem as described above.
MUs can achieve much better performance
by offloading their delay-sensitive or
computation-intensive tasks to the cloudlet nearby \cite{kumar2013survey},
because the cloudlets can provide them with
low-latency and rich computing resource access \cite{Jia2016Cloudlet}.

The resource allocation has been investigated in
the work \cite{2017ICC}, and the cloudlet deployment for task offloading  has been discussed
in  \cite{Xia2013Throughput}, \cite{Satyanarayanan2001Pervasive}.
Many efficient algorithms have been proposed in \cite{Jia2015Optimal}, \cite{Xu2015Capacitated}, to balance the workload among the cloudlets
for reducing the MUs' delay.
But access points (APs) and cloudlets may be reluctant
to provide those services without any rewards,
due to selfishness. To inspire cloudlets sharing their
resources with MUs, incentive mechanisms have been
introduced \cite{Jin2015Auction}, \cite{Samimi2014A}.
However, one cloudlet only serve one MU in those works.
Moreover, the resource in a cloudlet is always too expensive to be employed by a single MU.

To solve the above problems, there are several challenges: 1) How to place the cloudlets at APs efficiently. 2) How to assign cloudlet resources to the MUs when each MU has limited budget. 3) How to provide incentive for the three kinds entities (MUs, APs, Cloudlets).

Therefore, motivated by the group-buying scheme for spectrum allocation \cite{Lin2013Groupon}, we propose three efficient auction schemes to solve the problems of cloudlet placement and resource assignment jointly, which consists of three stages in each scheme.
In the first stage, we divide all MUs into several small groups of MUs
according to the AP they connected to, and then we
figure out the total budget for each group of MUs.
In the second stage, we assign cloudlets to APs.
Finally, we charge MUs in the third stage
according to the matching results.

The main contributions of this work can be summarized as follows.
\begin{enumerate}
\item We propose three auction schemes for joint cloudlet
      placement and resource assignment.
      The first scheme randomly generates
      a number $m$ according to the capacity of each  given cloudlet,
      followed by selecting the first $m$ MUs
      according to the performance price ratio, calculating
      the budget for the given cloudlet.
\item Based on the first scheme, the second scheme calculates several profitable cases and then randomly selects one from them. It can improve the revenue of the small MU groups significantly.
      In the third scheme, we match cloudlets with APs in a global way based on the second scheme.
\item We prove that all three schemes can work in
      polynomial time.
      We also provide proofs for individual rationality, budget balance and truthfulness.
      Both theoretical analysis and simulation results show that
      the proposed schemes outperform  the existing work
      in this paper.
\end{enumerate}

The rest of the paper is organized as follows.
Section \ref{sec:1} describes related works about
incentive mechanisms for resource allocation in mobile cloud computing.
Section \ref{sec:2} formulates the resource allocation problem and
 describe the three-stage auction model.
Section \ref{sec:3} introduces our algorithms
 in the auction model, together with some examples.
 In section \ref{sec:4}, we prove the economic properties
 for the proposed auction model.
 Simulation results are given in section \ref{sec:5}.
Finally, section \ref{sec:6} concludes this paper.

\section{Related Work}
\label{sec:1}
Resource allocation in mobile cloud computing is one of the
fresh and meaningful topics in recent years \cite{2016TMC_MCC}, \cite{2014_Survey_MCC}.
Mobile users  offload their heavy tasks to the
neighboring cloudlet, this has been an appealing
way to relieve their demand for resources \cite{2015IEEE_Magazine_cloudlet}, \cite{2017IEEE_cloud_computing}.
For cloudlet deployment, many existing works such as
 \cite{Jia2016Cloudlet}, \cite{Jia2015Optimal},
 \cite{Xu2015Capacitated} care about the cloudlet
placement in a given network, and most of them
focus on allocation cloudlet resource in a centralized manner.
Mike and his partners \cite{Jia2016Cloudlet} \cite{Jia2015Optimal}
discuss the challenge of cloudlet load balancing, and
they proposed a useful algorithm which is fast and
scalable to balance the workload for each cloudlet
in the wireless metropolitan area networks.
In \cite{Xu2015Capacitated}, how to place cloudlets
is  first considered to reduce the processing delay
for tasks while the resource of the cloudlet is limited.
Authors propose a heuristic algorithm
and an approximation algorithm to place cloudlets.
However, those works \cite{Jia2016Cloudlet} \cite{Jia2015Optimal} \cite{Xu2015Capacitated} do not take the cost of
cloudlets and APs into consideration.
Cloudlets and APs in this system may feel
reluctant to share their resource to
the mobile users  without any reward.

Incentive mechanisms which
take those costs into consideration
have been discussed in \cite{2015TWC}.
Resource allocation schemes in those works
are more flexible and intelligent. Also,
the resource holder and relay nodes are willing to serve users.
The auction schemes are wildly used in the study of computer science, the details can be seen in \cite{2011ACM_Survey_auction}, \cite{2013IEEE_Survey_auction}.
In \cite{2015TWC}, a cooperative bargaining
game-theoretic algorithm is addressed for resource
allocation in cognitive small cell networks.
However, one cloudlet can only serve one MU in those works.
The group-buying idea is introduced in
 \cite{Lin2013Groupon} and \cite{yang2014truthful}.
In \cite{Lin2013Groupon}, a group-buying auction
model is proposed to manage the spectrum redistribution, and
the problem of that a single buyer
cannot afford the whole spectrum is fixed.

In this paper, we introduce group-buying model
into cloudlet deployment, to divide independent
MUs into small groups based on the associated APs.
Therefore, MUs of each group can afford those expensive cloudlets,
and the cloudlet may share its resources
with MUs in a flexible and efficient way.
Different from our conference version \cite{zhou2017tacd}, we have added one more auction scheme in this work and we have extended the conference work to better present the main idea of the three stage auction scheme.
%the text, performance figures, theorems, and algorithms are newly generated in this journal version.

\section{System Model and Problem Formulation}
\label{sec:2}

\subsection{Problem Formulation}

The MU can be regarded as the buyer in our auction schemes.
The cloudlet is constituted by
resource-abundant devices, it is also the seller
in our auction schemes.
The AP is the access point of
the wireless network for MUs, and it also can
be placed with a cloudlet to improve mobile devices'
performance, so it is the auctioneer between MUs and cloudlets.

Assume that the number of cloudlets is $K$.
$C_{k}$ indicates the $k$th cloudlet. $Cap^{k}$ indicates
the resource capacity of $C_{k}$.
As defined as in \cite{Kang2015Incentive},
the cost function of cloudlet is

\begin{equation}
Cos(k) = c(k) \cdot w(k),
\end{equation}
where $c(k)$ is the cost factor of $C_{k}$,
and $w(k)$ is the workload brought by MUs' offloaded tasks.
In this paper, we try to make the cloudlet share its
resources to a suitable small group of MUs rather than just one MU.
To inspire cloudlets sharing their resources,
we define the  reserve price of $C_{k}$, denoted as $r_{k}$,

\begin{equation}\label{fomul:rk}
r_{k} = c(k) \cdot Cap^{k} + \delta.
\end{equation}
Where $C_{k}$ must be paid at
least $r_{k}$, no matter which  group of MUs finally wins
$C_{k}$.
Cloudlets in this paper may be heterogeneous,
we assume that their capacity and cost factor
may be different with each other, so their
reserve price will also be different.
While cloudlet $C_{k}$ joins in the auction scheme,
its total resource capacity $Cap^{k}$ and cost
factor $c(k)$ are fixed. $C_{k}$ cannot change them
during the whole auction. Then, $r_{k}$ is also fixed.
By the way, $C_{k}$ can adjust its reserve price
$r_{k}$ by changing its parameter $\delta$ after
a whole auction, such as increasing the value of $\delta$ if
its resource is over competitive in the market,
and decreasing the value of $\delta$ while the resource
is oversupplied, which will make $C_{k}$ benefit more
from the auction, but this feedback mechanism is
out of the scope this paper. Therefore, we assume $\delta = 0$ in this paper.

Assume that the number of AP is $n$ in the given network.
$a_{i}$ indicates the $i$th AP, and it is connected with $n_{i}$ MUs.
In this paper, MUs connect to the wireless network through AP.
Therefore, we can easily divide MUs into some groups base
on the connected AP by the MUs.
Each group of MUs can be assigned at most one cloudlet,
and if the group of MUs which connects with $a_{i}$
is assigned with cloudlet $C_{k}$, the MUs in the group
cannot request other cloudlet resource,
and the cloudlet $C_{k}$ can only serve for the MUs
in the group of $a_{i}$.
It is noteworthy that this is different with \cite{Jia2015Optimal},
where MU can request service from other cloudlets
if it's local AP do not have cloudlet or the assigned cloudlet is out of service.
In our auction schemes, APs are the auctioneer
who deals with the transaction between MUs and cloudlets.

For MUs that connected with the wireless network
through the $i$th AP, we call them the $i$th
group of MUs. Different groups have different amount of tasks to offload.
Let $m_{i}^{j}$ be the $j$th MU of the $i$th AP.
Its valuation for each cloudlet may be different.
The mobile user $m_{i}^{j}$ may give a higher valuation
for the cloudlet it preferred
(such as the cloudlets which have a good quality
of service to it). Then it will submit a much
higher bid on those cloudlets based on their valuation.
Instead, $m_{i}^{j}$ will submit a much lower bid
on the cloudlets which $m_{i}^{j}$ do not like.
Then, the valuations of $m_{i}^{j}$ on the $k$th
cloudlet $C_{k}$ is $v_{i}^{j}(k)$, which is private
information of $m_{i}^{j}$. The  budget
of $m_{i}^{j}$ for $C_{k}$ is $b_{i}^{j}(k)$,
which is public information, as this budget is the
bid that MU submits for cloudlets.
Namely, MUs' valuation for each cloudlet depends
on their preference of those cloudlets, and is known only by themselves.
Different MUs may produce different valuations on the same cloudlet,
according  to their different preferences.
Usually, in an auction schemes, the buyer bid
truthfully only if its budget equals its valuation.
For instance, MU $m_{i}^{j}$ bid truthfully on cloudlet
$C_{k}$ only if $b_{i}^{j}(k) = v_{i}^{j}(k)$.
But MUs' valuation for each cloudlet is unknown
to others, so the auction scheme must be truthful
enough to prevent MU benefit more by bidding
untruthfully, or the auction will bankrupt soon.

When the transactions are done after our
three-stage auctions, the winner MUs will pay for
the winner cloudlets and the connected APs,
the winner cloudlets will be placed on its matching AP
and serve for the small group of MUs connected by this AP.
For instance, if MUs in $a_{i}$ wins $C_{k}$,
$C_{k}$ will be placed on $a_{i}$, and then $C_{k}$
provides services to MUs in $a_{i}$.
Let $w_{i}$ be the winner set, which consists of
the winner MUs in the group of MUs in $a_{i}$.
Let $p_{i}^{j}$  be  clearing price of the MU $m_{i}^{j}$.

If $m_{i}^{j}$ is a winner, then $m_{i}^{j}$ will be
charged at $p_{i}^{j}$ after the auction.
For the case of that $m_{i}^{j}$ bid truthfully,
we define its utility $u_{i}^{j}$ as

\begin{equation}
u^{j}_{i} =\left \{
\begin{array}[l]{lcl}
v_{i}^{j}(k) - p_{i}^{j}   &       &{if\ m^{j}_{i}\in w_{i},}\\
0   &       &{otherwise,}\\
\end{array} \right.
\end{equation}
where $v_{i}^{j}(k)$ is the  valuation of  $m_{i}^{j}$
on the cloudlet $C_{k}$ it wins. This equation implies
the $m_{i}^{j}$ obtains the benefit from the auction.
Similarly, the winner set $W$ contains the winner APs.
If $a_{i}$ is a winner AP, its clearing price is $P_{i}$.
When $a_{i}$ bid truthfully, its utility $u_{i}$ is defined as

\begin{equation}
u_{i} =\left \{
\begin{array}[l]{lcl}
R_{i}^{k} - P_{i}   &       &{if\ a_{i}\in W,}\\
0   &       &{otherwise,}\\
\end{array} \right.
\end{equation}
where $R_{i}^{k}$ is the actual revenue
that $a_{i}$ calculates for its winner cloudlet $C_{k}$.
Let $W'$ be the set of winner cloudlets,
and $P^{k}$ be the clearing price of  $C_{k}$.
Its utility $u^{k}$ is defined as

\begin{equation}
u^{k} =\left \{
\begin{array}[l]{lcl}
P^{k} - r_{k}   &       &{if\ C_{k}\in W',}\\
0   &       &{otherwise.}\\
\end{array} \right.
\end{equation}

The social welfare can quantify the efficiency of
our auction schemes. Let $SW$ be the social welfare,
which means the total utility of all participants in the auction.
It is defined as

\begin{equation}
SW = \sum_{i = 1}^{n}\sum_{j = 1}^{n_{i}}u_{i}^{j} + \sum_{i = 1}^{n}u_{i} + \sum_{k = 1}^{K}u^{k}.
\end{equation}

\begin{table}[htbp]\scriptsize
\vskip -3mm
 \centering
 \caption{\label{table:symbols1}Symbols of Participant}
 \begin{tabular}{C{3.5cm}C{1cm}C{1cm}C{1.3cm}}
  \hline\noalign{\smallskip}
  \textbf{Definition} & \textbf{$C_{k}$}& \textbf{$a_{i}$}& \textbf{$m_{i}^{j}$} \\
  \noalign{\smallskip}\hline\noalign{\smallskip}
  Quantity  &   $K$&    $n$&    $n_{i}$ for $a_{i}$\\
  Capacity or Workload&  $Cap^{k}$&$-$&$l_{i}^{j}$\\
  Cost, Revenue or Valuation&    $Cos(k)$ &   $R_{i}^{k}$&$v_{i}^{j}(k)$\\
  Reserve price or Budget&$r_{k}$&$B_{i}^{k}$&$b_{i}^{j}(k)$\\
  Clearing price& $P^{k}$  & $P_{i}$  & $p_{i}^{j}$\\
  Utility& $u^{k}$ & $u_{i}$  & $u_{i}^{j}$\\
  Winner set& $W'$  & $W$  & $w_{i}$\\
  \noalign{\smallskip}\hline
 \end{tabular}
 \vskip -3mm
\end{table}

\subsection{System Model}

\begin{figure}[h]
\vskip -3mm
	\centering
	\setlength{\belowcaptionskip}{-1em}
	\includegraphics[width=4.5in]{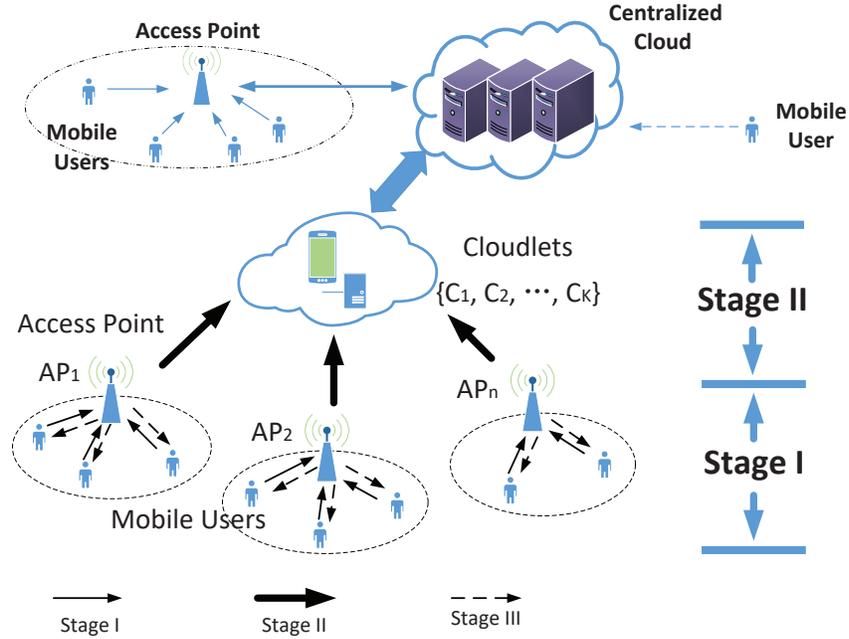}
	\caption{\label{fig:model} {\small Three-stage auction model.}}
\vskip -3mm
\end{figure}

The Fig. \ref{fig:model} shows the model of our
three-stage auction schemes.
In the first stage, we divide MUs into $n$ small groups
according to the APs that connect the MUs and cloudlets.
Then, in each group the AP calculates its total
revenue for each cloudlet, e.g. the AP $a_{i}$
calculates the revenue $R_{i}^{k}$ for cloudlet
$C_{k}$. $R_{i}^{k}$ is calculated according
to the budget of the MU group in $a_{i}$,
and these budgets are their bids for $C_{k}$,
i.e., $b_{i}^{j}(k) (j \in [1, \ldots, n_{i}])$.
The total revenue quantify the preference
of the MU group on each cloudlet.
In the AP $a_{i}$, the MU $m_{i}^{j}$ which
has been utilized in calculating $R_{i}^{k}$
will be regarded as a potential winner for
cloudlet $C_{k}$, and its potential price
is $p_{i}^{j}(k)$.
If $a_{i}$ wins $C_{k}$ in the next stage,
$C_{k}$ will share its resources with
$m_{i}^{j}$, and $m_{i}^{j}$ will be charged
at $p_{i}^{j}(k)$, i.e., its clearing price $p_{i}^{j}$
equals to $p_{i}^{j}(k)$.
On the other hand, $C_{k}$ will only share
its resources with the MU who paid for it.
We cannot ensure that all MUs in $a_{i}$ can be served
by $C_{k}$, due to the constraints of economic properties.
The rest of MUs will be left to the next round of the
auction, which is not within the scope of this paper.

In the second stage, APs submit their budget to each cloudlet.
This budget is the total budget of the MU group in the corresponding AP,
which is generated base on the revenue for each cloudlet.
For instance, the  budget of $a_{i}$ for $C_{k}$ is
$B_{i}^{k}$ which is the price that $a_{i}$ bid for $C_{k}$.
For each AP, its revenue $R_{i}^{k}$ is provided by its MU group.
It is a real value, and the budget $B_{i}^{k}$ is generated
by itself, we can easily find that both $R_{i}^{k}$ and $B_{i}^{k}$
are public information. Therefore, we can easily verify
that whether $a_{i}$ bid truthfully or not.
After that, we try to match cloudlets with APs while subjecting to our desired properties.
As a result, for the winner set of cloudlets $W'$
and the winner set of APs $W$, the matching result
between $W'$ and $W$ can be defined by the mapping function $\sigma(\dot)$.
For example, $\sigma(i) = k$ means cloudlet $C_{k}$ is
assigned to AP $a_{i}$, and their clearing prices
$P_{i}$ and $P^{k}$ are same.

Then, in the third stage, the winner MUs set in
$a_{i}$ is $w_{i}$, winning APs will charge them according to their potential winner
price generated in the first stage.

\subsection{Desirable Properties}

\subsubsection{\quad \quad \quad  Truthfulness}

Let $\theta$ be a positive number.
The participants may pay an extra cost $\theta$ to figure out
how to bid in the auction scheme that makes them benefit more.
When MU $m_{i}^{j}$ bid untruthfully, we
define the utility $\widetilde{u}_{i}^{j}$  as follows.

\begin{equation}
\widetilde{u}_{i}^{j} =\left \{
\begin{array}[l]{lcl}
v_{i}^{j}(k) - p_{i}^{j} - \theta  &       &{if\ m^{j}_{i}\in w_{i},}\\
-\theta   &       &{otherwise.}\\
\end{array} \right.
\end{equation}

Similarly, we now define the utility $\widetilde{u}_{i}$
for the case of that AP $a_{i}$ bid untruthfully.

\begin{equation}
\widetilde{u}_{i} =\left \{
\begin{array}[l]{lcl}
R_{i}^{k} - P_{i} - \theta  &       &{if\ a_{i}\in W,}\\
-\theta   &       &{otherwise.}\\
\end{array} \right.
\end{equation}

The extra cost $\theta$ varies for different MUs
and different APs. The different market situation
also causes different extra cost even for the same MU (or the same AP).
In this paper, we define truthfulness as a weakly
dominate strategy as mentioned in \cite{Jin2015Auction},
where the player cannot improve its utility
by bidding an untruthful bid in truthful auction scheme.
Truthfulness is significant for an auction,
we must ensure $u_{i}^{j} \geq \widetilde{u}_{i}^{j}$
and $u_{i} \geq \widetilde{u}_{i}$ for each MUs and APs
to keep our auction truthful.
In our auction scheme, we discuss the truthfulness
in which only one player can change its bid or strategy,
and the others cannot.

\subsubsection{\quad\quad \quad Budget balance}

The total price  charging for buyers is not
less than the total price paid for sellers.
If $\sigma(i) = k$, then
  $$\sum_{i=1}^{n}\sum_{j=1}^{n_{i}}p_{i}^{j} \geq \sum_{k=1}^{K}P^{k} + (\sum_{i=1}^{n}R_{i}^{k} - \sum_{i=1}^{n}P_{i}).$$

\subsubsection{\quad\quad \quad Individual rationality}

For sellers, they cannot benefit at a price
smaller than it's asked, i.e., $P^{k} \geq r_{k}$.
For buyers, they cannot be charged at a
price bigger than it's bid, i.e., $b_{i}^{j}(k) \geq p_{i}^{j}(k) = p_{i}^{j}$ if $\sigma(i) = k$.
For APs, we define their individual rationality as
$R_{i}^{k} \geq B_{i}^{k} \geq P_{i}$ if $\sigma(i) = k$.

\subsubsection{\quad\quad \quad Computation efficiency}

We will prove that the schemes can be performed in polynomial time.

\section{Auction Schemes}

\label{sec:3}

In this section, we describe the proposed three auction schemes.
The first is  for Three-stage Auction scheme for Cloudlet
Deployment, named TACD. The second, named TACDp, is an improved
version of TACD by refining the first stage of the auction scheme.
The third is called TACDpp, that is derived from TACDp by
improving the mapping approach in its second stage.

\subsection{Framework of the Schemes}

All these three schemes are inspired by the idea of
``group-buying''. Each scheme consists of three stages.
In stage \uppercase\expandafter{\romannumeral1},
APs calculate the revenue from their small group of MUs,
and figure out the potential winner MUs for each cloudlet,
the algorithm used in this stage is named ACRC.
The revenue matrix is indicated as
$\{R_{i}^{k}\} (i \in [1, \ldots, n], k \in [i, \ldots, K])$,
which is formed by the revenues of the APs for each cloudlet.
APs can bid for cloudlets according to $\{R_{i}^{k}\}$,
and these bids form the budget matrix
$\{B_{i}^{k}\} (i \in [1, \ldots, n], k \in [i, \ldots, K])$.
In stage \uppercase\expandafter{\romannumeral2},
we match APs with cloudlets according to the budget
matrix $\{B_{i}^{k}\}$ and the reserve price vector
$\{r_{k}\} (k \in [1, \ldots, K])$,
where the vector is formed by the reserve price of
cloudlets, and the algorithm in this stage named ASC.
In stage \uppercase\expandafter{\romannumeral3}, the
winner APs, which are placed with cloudlets, allocate
resources to their winner MUs and charge these MUs.

\begin{table}[htbp]\scriptsize
\vskip -3mm
 \centering
 \caption{\label{table:symbols2}Symbols in Algorithms}
 \begin{tabular}{cc}
  \hline\noalign{\smallskip}
  \textbf{Symbol} & \textbf{Definition } \\
  \noalign{\smallskip}\hline\noalign{\smallskip}

$t_{i}^{j}(k)$ &$m_{i}^{j}$'s performance price ratio on $C_{k}$\\
$A$ &Array of MU sorted by $t_{i}^{j}(k)$\\
$l_{x}$&The workload of the $x$th MU in $A$\\
$A_{x}$, ${L_{x}}$&The first $x$ MUs in $A$, and their total workload \\
$s$&The maximum quantity of MUs in $A$ while $L_{s} \leq Cap^{k}$\\
$S_{x}$&The revenue of the first $x$ MUs in $A$\\
$m$&The independent integer\\
$w_{i}^{k}$&The potential winner MUs in $a_{i}$ for $C_{k}$\\
$p$&The unit price of MUs\\
$p_{i}^{j}(k)$&$m_{i}^{j}$'s potential price on $C_{k}$\\
$top_1, top_2$&The top factor in ACRC, ASC\\
$A'$&The randomly sorted AP set\\
$D$&The profit matrix $\{B_{i}^{k} - r_{k}\}$\\
%$top2$&The top $top2$ valuable profit in $D$ \\
$\sigma$&Mapping function from $a_{i}$ to $C_{k}$\\
%$U^{k}$&$C_{k}$'s utilization\\
 \noalign{\smallskip}\hline

 \end{tabular}
\vskip -3mm
\end{table}

%require = input   ensure = output
\renewcommand{\algorithmicrequire}{\textbf{Input:}}
\renewcommand{\algorithmicensure}{\textbf{Output:}}

\vskip -3mm
\begin{algorithm}[h]
    \small
	\begin{algorithmic}[1]
		\caption{ACRC: AP $a_{i}$ Calculating the Revenue vector for each Cloudlet}\label{alg:ACRC}
    \REQUIRE{Sorted MUs array $A$,  cloudlets' capacity set $\{ Cap^{k} \}$}
    \ENSURE{$a_{i}$'s revenue vector $\{ R^{k}_{i} \}$, $a_{i}$'s potential winner matrix $\{ w_{i}^{k} \}$ and $a_{i}$'s potential price matrix $\{p_{i}^{j}(k)\}$}
    \FOR{$k = 1$ to $K$ }
%        \STATE{Let $A$ be MUs array sorted by their performance price ratio \{$t^{j}_{i}(k)$\} in descending order.}
%        \STATE{Let $l_{s}$ be the $s$th MU's workload in $A$.}
%        \STATE{$L_{s}$ means the total workload of the first $s$ MUs in $A$, \\ $L_{s} = l_{1} + l_{2}+...+l_{s}$.}
        \STATE{Maximizing the number $s$ subject to $L_{s} \leq Cap^{k}$ and $L_{s} + l_{s+1} > Cap^{k}$.}
        \STATE{If $L_{n_{i}} \leq Cap^{k}$, then $s = n_{i}$.}
        \STATE{The revenue set $\{S_{x}\} = GTR(A, s)$, and the revenue of the first $s-1$ cases is $S_{1}, S_{2}, \ldots, S_{s-1}$.}
        \STATE{The integer $m$ is randomly generated in $[(s+1)/2, s-1]$.}
        \STATE{Then the revenue $R_{i}^{k} = S_{m}$.}
        \STATE{$a_{i}$'s potential winner set for $C_{k}$ is $w_{i}^{k} = A_{m}$. }
        \STATE{Then the unit price $p$ equals to the $(m+1)$th MU's performance price ratio in $A$.}
        \IF{$m_{i}^{j} \in w_{i}^{k}$}
        \STATE{$p_{i}^{j}(k) = l_{i}^{j} \cdot p$}
        \ELSE
        \STATE{$p_{i}^{j}(k) = 0$}
        \ENDIF
    \ENDFOR
    \RETURN{$\{ R_{i}^{k} \}$, $\{ w_{i}^{k} \}$, $\{p_{i}^{j}(k)\}$}
\end{algorithmic}	
\end{algorithm}
\vskip -3mm

\begin{algorithm}[h]
    \small
	\begin{algorithmic}[1]
		\caption{GTR: Getting the Revenue set}\label{alg:GTR}
    \REQUIRE{$A$, $s$}
    \ENSURE{The revenue set $\{S_{x}\}$}
    \STATE{Let $\{S_{x}\}$ be the revenue set of the first $s-1$ cases in $A$. }
    \FOR{$x = 1$ to $s-1$}
        \STATE{The unit price $p$ is equal to the $(x+1)$-th MU's performance price ratio in $A$.}
        \STATE{$L_{x}$ is total workload of the first $x$ MUs in $A$.}
        \STATE{Then $S_{x} = p \cdot L_{x}$.}
    \ENDFOR
    \RETURN{$\{S_{x}\}$}
    \end{algorithmic}	
\end{algorithm}
\vskip -3mm

\subsection{ Scheme 1: TACD}

\subsubsection{\quad\quad \quad Stage \uppercase\expandafter{\romannumeral1}: Calculating Revenue}

The algorithm used in the first stage of TACD is named ACRC.
For more details, see Algorithm \ref{alg:ACRC}.
At first, for each AP such as $a_{i}$, we calculate
its revenue $R_{i}^{k}$ for all cloudlets.
Obviously, the revenue $R_{i}^{k}$ is calculated
from the small group of MUs in $a_{i}$.
Let $t_{i}^{j}(k)$ be the performance price ratio
of the MU $m_{i}^{j}$.

In other words,  $t_{i}^{j}(k)$ is the  unit
budget of $m_{i}^{j}$ for the cloudlet $C_{k}$,
and it is defined as follows.

\begin{equation}
t_{i}^{j}(k) = \frac{b_{i}^{j}(k)}{l_{i}^{j}},
\end{equation}
where $l_{i}^{j}$ is the workload of $m_{i}^{j}$,
and the value of $l_{i}^{j}$ is kept unchange
no matter which cloudlet receives the tasks offloaded by $m_{i}^{j}$.
The value of $t_{i}^{j}(k)$ will increase with the increasing
$b_{i}^{j}(k)$, i.e., $m_{i}^{j}$ will get a higher
performance price ratio on $C_{k}$ if it has more budget on $C_{k}$.

The set $A$ consists of the MUs in $a_{i}$,
where the MUs are sorted in descending order
in terms of their performance price ratio $t^{j}_{i}(k)$.
%For example, assume $A = \{m_{i}^{a}, m_{i}^{b}, m_{i}^{c}, ...\}$.
%Then, we have $t_{i}^{a}(k) \geq t_{i}^{b}(k) \geq t_{i}^{c}(k) \geq ...$.
Let $A_{x}$ be  the set of  MUs  which are the
first $x$ ($x \leq n_{i}$) members of $A$.
%e.g., $A_{3} = \{m_{i}^{a}, m_{i}^{b}, m_{i}^{c}\}$ and $A_{n_{i}} = A$.
Let $l_{x}$ be the workload of the $x$th MU in $A$,
i.e., $l_{1}$ is the workload of the first MU in $A$.
Let $L_{x}$ be the total workload of $A_{x}$,
i.e., $L_{x} = l_{1}+l_{2}+l_{3}+...+l_{x}$.
We try to find the index $s$ in $A$ to maximize
$L_{s}$, in which $L_{s} \leq Cap^{k}$ and $L_{s} + l_{s+1} > Cap^{k}$.
If the total workload of the MUs in $a_{i}$
is less than or equal to $Cap^{k}$, i.e.,
$L_{n_{i}} \leq Cap^{k}$, then $s = n_{i}$.

Let $S_{x}$ be the revenue which is generated
by the first $x$ MUs of $A$.
Let $S_{x} = p \cdot L_{x}$, where $p$ is the
unit price which equals to the performance price
ratio of the $(x+1)$th member in $A$.
The Algorithm \ref{alg:GTR} which named GTR is to
get the revenue set $\{S_{x}\}$, where
$\{S_{x}\} = \{S_{1}, S_{2}, \ldots, S_{s-1}\}$.
In order to keep the MUs bid truthfully, we randomly
generate an integer $m$, where $(s+1)/2 \leq $m$ \leq s-1 $.
The random number $m$ is independent of  the bids of MUs'.
Then $a_{i}$'s revenue for $C_{k}$ equals to $S_{m}$,
i.e., $R_{i}^{k} = S_{m}$.
The set of potential winner of $a_{i}$  for $C_{k}$
consists of the first $m$ MUs of $A$, i.e., $w_{i}^{k} = A_{m}$.
The unit price $p$ equals to the performance price ratio
of the $(m+1)$th MU in $A$.
For the MU $m_{i}^{j}$ in $a_{i}$, its potential price
on $C_{k}$ is $p_{i}^{j}(k)$, and $p_{i}^{j}(k) = l_{i}^{j} \cdot p$ if $m_{i}^{j} \in w_{i}^{k}$, or $p_{i}^{j}(k) = 0$ if $m_{i}^{j} \notin w_{i}^{k}$.
It means if $a_{i}$ is allocated with $C_{k}$
after the whole auction scheme, then the MUs which
$m_{i}^{j} \in w_{i}^{k}$ are winners, and they will
be charged at $p_{i}^{j}(k)$ by $a_{i}$.
The sum of $\{p_{i}^{j}(k)\}$ equals to $R_{i}^{k}$,
i.e., $R_{i}^{k} = \Sigma_{j = 1}^{n_{i}}p_{i}^{j}(k)$,
that is the preference of the MUs in $a_{i}$ for the cloudlet $C_{k}$.

In TACD, we choose the random number $m$ in
$[s+1)/2, s-1]$ based on the following reasons.
First, the number $m$ must be a random number
to keep our auction truthful, and we will discuss it later.
Second, if the random number $m$ is close to $1$,
the unit price $p$ will be increased but the number
of winner MUs will be reduced, and it will go opposite side
 if $m$ close to $s$.
The performance comparisons for different $m$ values
are mentioned in \cite{Lin2013Groupon}, the authors addressed
that the APs  will get more budget while the number of MUs
fall in $[30\%, 70\%]$.
Similarly, in this paper the APs will get more budget
when the number $m$ is randomly generated in $[s+1)/2, s-1]$.
Third, for each AP, the more budget it calculates
the easier it wins a more profitable cloudlet
in the second stage.
Finally, if AP gets the same revenue
at $m_{1} = 0.3s$ and $m_{2} = 0.7s$, it will win the next
stage at the same probability, but there is a big
difference between  the social welfare derived from the two settings of
$m$. It is clear that $m = 0.7s$ is better.
In summary, we generate the random number in
$[s+1)/2, s-1]$, so that AP can calculate a higher
budget and get more profits.

\begin{figure}[h]
\vskip -3mm
	\centering
	\setlength{\belowcaptionskip}{-1em}
	\includegraphics[width=3.5in]{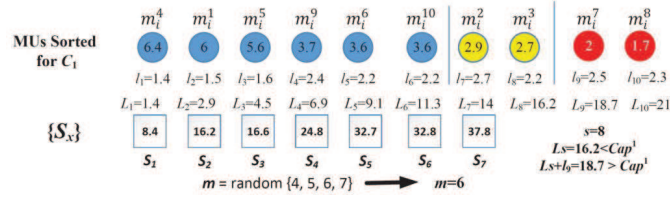}
	\caption{\label{fig:example1} {\small Illustration of ACRC in TACD.}}
\vskip -3mm
\end{figure}

To illustrate the detail of ACRC in TACD,
we provide an simple example to demonstrate how this
algorithm works for AP $a_{i}$.
In this example, the  performance price ratios of the MUs
on $C_{1}$ and $C_{2}$ are shown in Table \ref{table:example1}(a).
Their workload vector is shown in Table \ref{table:example1}(b),
and the capacity vector of cloudlet  is shown in
Table \ref{table:example1}(c).
For cloudlet $C_{1}$, we sort MUs in terms of  their
performance price ratio $t_{i}^{j}(1)$
in descending order at first.
Then the order of MUs in the sorted array $A$ is:
$A = \{ m_{i}^{4}, m_{i}^{1}, m_{i}^{5}, m_{i}^{9}, m_{i}^{6}, m_{i}^{10}, m_{i}^{2}, m_{i}^{3}, m_{i}^{7}, m_{i}^{8}\}$.
Let $l_{s}$ be the workload of the $s$th MU in $A$.
The workloads of the MUs in $A$ are
$\{ l_{1}=1.4, l_2=1.5, l_{3}=1.6, ... \}$,
which are shown in Fig. \ref{fig:example1}.
Let $L_{x}$ be the total workload of the first $x$
members of $A$. For instance, $L_{3} = l_{1}+l_{2}+l_{3} = 4.5$.
According to ACRC, $s = 8$, MUs $m_{i}^{7}$
and $m_{i}^{8}$ which are painted in red are losers in ACRC.
Then we calculate the revenue for this $s-1$ cases.
The unit price $p$ for $S_{x}$ is the $(x+1)$-th
performance price ratio of MU in $A$, and
$S_{x} = p \cdot L_{x}$. For instance, the unit price
$p$ for $S_{5}$ is the $6$th  performance price ratio of
MU in $A$, i.e., $p = t_{i}^{10}(1) = 3.6$.
Then, $S_{5} = p \cdot L_{5} = 3.6 * 9.1 = 32.76$.
We get a random number within $( 4, 7 )$.  Assume that $m =6$.
We `sacrifice' MUs $m_{i}^{2}, m_{i}^{3}$ which are
painted in yellow to keep ACRC truthful.
Therefore $R_{i}^{1} = S_{6} = 32.8$ and the unit price
$p = t_{i}^{2}(1) = 2.9$. The first $6$ MUs in this
example form the potential winner set for $C_{1}$,
i.e., $w_{i}^{1} = \{ m_{i}^{4}, m_{i}^{1}, m_{i}^{5}, m_{i}^{9}, m_{i}^{6}, m_{i}^{10} \}$.
For these MUs, their potential price $p_{i}^{j}(k) = p \cdot l_{i}^{j}$.
In this example, their potential price in $A$ is
$p_{i}^{4}(1) = 2.9 * 1.4 = 4.06$, $p_{i}^{1}(1) = 2.9 * 1.5 = 4.35$, $p_{i}^{5}(1) = 2.9 * 1.6 = 4.64$, and $p_{i}^{9}(1) = 6.96$, $p_{i}^{6}(1) = 6.38$, $p_{i}^{10}(1) = 6.38$.
For the rest of  MUs $m_{i}^{j} \notin w_{i}^{1}$,
their potential price $p_{i}^{j}(1) = 0$.
Then we can get the potential price set $\{p_{i}^{j}(1)\}$.
It is similar for $C_{1}$ when  AP $a_{i}$ calculates revenue for other cloudlets.

After all the APs have calculated the revenue of
each cloudlet, the revenue matrix $\{R_{i}^{k}\}$ is formed.
Then APs will bid for each cloudlet in the next stage.
These bids constitute the budget matrix $\{B_{i}^{k}\}$
which means the APs' budget for each cloudlet.
All these APs have submitted their truthful bid
if $\{B_{i}^{k}\} = \{R_{i}^{k}\}$, or there must be one/some
cheater(s). The later case is what we need to avoid.

\begin{table}
	\centering{
\caption{Example for ACRC}\label{table:example1}
	\begin{subtable}{(a) MUs' performance price ratio on each Cloudlet}
		\centering
		 \begin{tabular}{cccccc}
          \hline\noalign{\smallskip}
           &$t_{i}^{1}(k)$&$t_{i}^{2}(k)$&$t_{i}^{3}(k)$&$t_{i}^{4}(k)$&$t_{i}^{5}(k)$ \\
          \noalign{\smallskip}\hline\noalign{\smallskip}
            $C_{1}$& 6 & 2.9&2.7&6.4&5.6 \\
            $C_{2}$& 6 & 2.5&4.5&5.7&3.1 \\
            ...&  & &&& \\
          \noalign{\smallskip}\hline\noalign{\smallskip}
            &$t_{i}^{6}(k)$&$t_{i}^{7}(k)$&$t_{i}^{8}(k)$&$t_{i}^{9}(k)$&$t_{i}^{10}(k)$ \\
          \noalign{\smallskip}\hline\noalign{\smallskip}
            $C_{1}$& 3.6 & 2&1.7&3.7&3.6 \\
            $C_{2}$& 1.8 & 3.2&4.3&3.7&2.9 \\
            ...&  & &&& \\
          \noalign{\smallskip}\hline
         \end{tabular}
		\label{table:example1.1}
	\end{subtable}
}
	\begin{subtable}{(b) The total workload of MUs' offloading task(s)}
		\centering
		\begin{tabular}{cccccccccc}
          \hline\noalign{\smallskip}
           $l_{i}^{1}$&$l_{i}^{2}$&$l_{i}^{3}$&$l_{i}^{4}$&$l_{i}^{5}$&$l_{i}^{6}$&$l_{i}^{7}$&$l_{i}^{8}$&$l_{i}^{9}$&$l_{i}^{10}$ \\
          \noalign{\smallskip}\hline\noalign{\smallskip}
            1.5&2.7&2.2&1.4&1.6&2.2&2.5&2.3&2.4&2.2 \\
          \noalign{\smallskip}\hline
         \end{tabular}
        \label{table:example1.2}
	\end{subtable}
    \begin{subtable}{(c) Cloudlets' resource capacity}
		\centering
		\begin{tabular}{cccccccc}
          \hline\noalign{\smallskip}
           $Cap^{1}$&$Cap^{2}$&$Cap^{3}$&$Cap^{4}$&$Cap^{5}$&$Cap^{6}$&$Cap^{7}$&...\\
          \noalign{\smallskip}\hline\noalign{\smallskip}
            17&22&25&11&19&21&18&... \\
          \noalign{\smallskip}\hline
         \end{tabular}
        \label{table:example1.3}
	\end{subtable}
\end{table}

\subsubsection{\quad\quad \quad Stage \uppercase\expandafter{\romannumeral2}: Matching Cloudlet for AP}

The algorithm used in this stage is named ASC,  more details are shown
in Algorithm \ref{alg:ASC}. In this stage, APs deal with cloudlets
according to the budget of APs and the reserve price of cloudlets.
In TACD, we assign cloudlet to AP in a greedy manner,
as mentioned in the existing work \cite{Goldberg2010Competitive}.
In ASC, we generate the profit matrix $D$ at first,
where $D = \{B_{i}^{k}\} - \{r_{k}\}$, and $d_{i}^{k} = B_{i}^{k} - r_{k}$.
Then we distribute the terms of APs randomly to $A'$.
For each AP in $A'$, we try to match it with an available
cloudlet $C_{k}$ to maximize the profit $B_{i}^{k}-r_{k}$ by the algorithm FRM.

The algorithm FRM is shown in Algorithm \ref{alg:FRM}.
For the AP $a_{i}$ in $A'$, then we try to match it with
a most profitable cloudlet among the rest of available cloudlets.
The profit vector of $a_{i}$ is $D_{i}$ which is the
$i$th row of the matrix $D$.
Then we select the largest element $d_{i}^{k}$ in  $D_{i}$.
The cloudlet $C_{k}$ is the most profitable cloudlet
for $a_{i}$ among the rest of available cloudlets.
If ties, we choose the $C_{k}$ with the smaller $k$.
As a result, FRM matches $a_{i}$ with $C_{k}$ and return the matching to ASC.
For this AP-cloudlet matching, its profit is $d_{i}^{k}$.
Then, the algorithm ASC judges that if their profit
is a positive value, i.e., whether $d_{i}^{k} > 0$.
The budget of $a_{i}$ is bigger than the reserve
price of $C_{k}$ if $d_{i}^{k} > 0$, i.e., if $B_{i}^{k} > r_{k}$.
Then we try to find a bid for $C_{k}$ from the
other APs. The selected bid has the biggest
value between $B_{i}^{k}$ and $r_{r}$.
In other words, we try to find the $B_{j}^{k}$
where $B_{i}^{k} \geq B_{j}^{k} \geq \ldots \geq r_{k}$ and $i \neq j$.
If there is no such $B_{j}^{k}$, then $a_{i}$ fails to
be allocated with $C_{k}$, and we set $d_{i}^{k} = 0$.
Otherwise, we allocate $C_{k}$ on $a_{i}$, i.e., let $\sigma(i) = k$.
The clearing prices of $a_{i}$ and $C_{k}$ equal
to the highest bid between $B_{i}^{k}$ and $r_{k}$,
i.e., $P_{i} = P^{k} = B_{j}^{k}$.
Then we add $a_{i}$ and $C_{k}$ in their winner set,
such as $W = W \cup a_{i}$ and $W' = W' \cup C_{k}$.
Finally, for the matrix $D$ we set the values of
all elements in the $i$th row to $0$. Meanwhile,
we also set the values of
all elements in the $k$th column to $0$.

Algorithm ASC can ensure the utility of both APs and
cloudlets if they are winners in the auction.
For each winner AP-cloudlet matching, their clearing
price $P_{i}, P^{k}$ are independent with $B_{i}^{k}$ and $r_{k}$,
both $a_{i}$ and $C_{k}$ cannot modify the clearing price by themselves.
This is helpful to keep the auction truthful.

\vskip -3mm
\begin{algorithm}
\small
	\begin{algorithmic}[1]
		\caption{ASC: APs' auction to Select suitable Cloudlet}\label{alg:ASC}
    \REQUIRE{$\{ B_{i}^{k} \}$, $\{ r_{k} \}$, $D$}
    \ENSURE{ $W$, $W'$, $\{P_{i}\}$, $\{P^{k}\}$, $\sigma$}
    \STATE{Distributing APs randomly into $A'$.}
    \FOR{$x = 1$ to $n$}
        \STATE{Getting AP $a_{i}$ and its matching cloudlet $C_{k}$ by algorithm FRM($D$, $A'$, $x$).}
        \IF{$d_{i}^{k} > 0$}
%            \IF{There is/are other AP's/APs' bid(s) is less or equal to $B_{i}^{k}$ and greater or qeual to $r_{k}$, i.e., }
            \IF{$B_{i}^{k} \geq B_{j}^{k} \geq \ldots \geq r_{k}$, which $j \neq i$}
            \STATE{$\sigma(i) = k$}
            \STATE{$P_{i} = P^{k} = B_{j}^{k}$}
            \STATE{$W = W \cup a_{i}$}
            \STATE{$W' = W' \cup C_{k}$}
            \STATE{Setting the values of elements in $i$th row of $D$ to $0$}
            \STATE{Setting the values of elements in $k$th column of $D$ to $0$}
            \ELSE
            \STATE{$d_{i}^{k} = 0$}
            \ENDIF
        \ENDIF

    \ENDFOR
    \RETURN{$W$, $W'$, $\{P_{i}\}$, $\{P^{k}\}$, $\sigma$}
\end{algorithmic}	
\end{algorithm}

\begin{algorithm}
    \small
	\begin{algorithmic}[1]
		\caption{FRM: Finding a Rational Matching to $a_{i}$}\label{alg:FRM}
    \REQUIRE{$D$, $A'$, $x$}
    \ENSURE{AP $a_{i}$ and it's matching cloudlet $C_{k}$}
    \STATE{Let $a_{i}$ denote the $x$th AP of $A'$.}
    \STATE{Let vector $D_{i}$ be the $i$th row of matrix $D$.}
    \STATE{$d_{i}^{k}$ is the maximum of $D_{i}$.}
    \RETURN{$a_{i}$, $C_{k}$}
    \end{algorithmic}	
\end{algorithm}

\subsubsection{\quad \quad  \quad Stage III: Charging for winner}

In this stage, the winner APs choose the winner MUs
according to their potential winner set, and then charge
them at their potential winner price $p_{i}^{j}(k)$.
For instance, while $a_{i}$ wins $C_{k}$ in stage
\uppercase\expandafter{\romannumeral2}, the MUs in the
potential winner set $w_{i}^{k}$ is the winner MUs of $a_{i}$.
For each MU $m_{i}^{j}$ where $m_{i}^{j} \in w_{i}^{k}$,
it will be charged by $a_{i}$ at the clearing price
$p_{i}^{j}$, where $p_{i}^{j} = p_{i}^{j}(k)$.

\subsection{Scheme 2: TACDp}

In this subsection, we propose a more efficient
scheme named TACD plus (TACDp).
The TACDp improves the first stage of TACD by
changing the generation method of $m$ in ACRC,
so that the APs in TACDp can get more revenue.
In TACD, $m$ is randomly generated in $[(s+1)/2, s-1]$,
it may sacrifice many MUs, resulting in the
performance decrease of TACD, although  it can
keep the auction scheme truthful.
In TACDp, we calculate several profitable revenues
and then randomly select one from them as the revenue
of the target AP.  In this section, we assume that the
default value of $top_1$ is $3$.
Then, we select the top $3$ profitable revenues
$S_{x_{1}}, S_{x_{2}}, S_{x_{3}}$
from $S$, and $m$ is randomly selected from $\{ x_1, x_2, x_3\}$,
denoted as $m = random\{x_1, x_2, x_3\}$.
We can also change the value of $top_1$ to get a better
result, e.g., $top_1 = 2$, then we select the top $2$ profitable
revenues $S_{x_{1}}, S_{x_{2}}$ from $S$,
and $m = random\{x_1, x_2\}$. The different value of $top_1$
will lead to different average revenue
and different degree of truthfulness. The effect of $top_1$
will be discussed in the next section.

\begin{figure}[h]
	%\centering
	\setlength{\belowcaptionskip}{-1em}
	\includegraphics[width=4.5in]{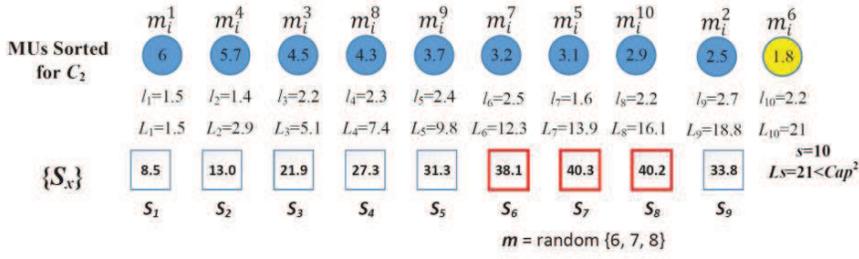}
	\caption{\label{fig:example2} {\small Illustration of ACRC in TACDp.}}
\end{figure}

To illustrate the first stage of TACDp, we calculate the
revenue of $a_{i}$ on $C_{2}$, which is shown in Table \ref{table:example1}.
The ACRC in TACDp is shown in Fig. \ref{fig:example2}.
In this example, $top_1 = 3$. Following TACD, we generate the number $s$ and
the revenue set $S$, resulting in $s = 10$ and
$S = \{8.5, 13.0, 21.9, 27.3, 31.3, 38.1, 40.3, 40.2, 33.8\}$.
The top $3$ cases in $S$ is $S_{7}, S_{8}, S_{6}$,
then $m =$ $random$ $\{6, 7, 8\}$, the average revenue is $39.5$.
It is worthwhile to point out that, the average revenue in TACD is $36.7$.
Thus, the revenue of the APs in TACDp is improved.

The rest steps of TACDp are the same with TACD.
Note that, the value of $top_1$ must
be larger than $1$. In this case, let $S_{max}$ be the
most profitable revenue of $S$, we cannot fix the
revenue of AP at $S_{max}$. This is because,
we cannot keep ACRC truthful if we always choose $R_{i}^{k} = S_{max}$.
For instance, $S_{max} = S_{7}$, i.e.,$40.3$ in Fig \ref{fig:example2}
and the unit price $p = t_{i}^{10}(2)$, i.e., $2.9$ while MUs bid truthfully.
For $m_{i}^{5}$, it's valuation on $C_{2}$ is
$v_{i}^{5}(2)$ where $v_{i}^{5}(2) = b_{i}^{5}(2)$,
i.e., $ 4.96$, its potential price
$p_{i}^{5}(2) = p \cdot l_{i}^{5} = 2.9 * 1.6 = 4.64$.
We assume that $a_{i}$ wins $C_{2}$ in
stage \uppercase\expandafter{\romannumeral2},
then $m_{i}^{5}$ will be charged at clearing
price $p_{i}^{5} = p_{i}^{5}(2) = 4.64$ in
stage \uppercase\expandafter{\romannumeral3}.
Therefore, the utility of $m_{i}^{5}$ is $u_{i}^{5}$
where $u_{i}^{5} = v_{i}^{5}(2) - p_{i}^{5} = 4.96 - 4.64 = 0.32$.
However, if $m_{i}^{5}$ bid untruthfully, we assume
that $m_{i}^{5}$ changes its budget on $C_{2}$ to
$b_{i}^{5}(2) = 4.32$ which is less than its
valuation on $C_{2}$.
Then the performance price ratio of $m_{i}^{5}$
on $C_{2}$ is $t_{i}^{5}(2)$ where $t_{i}^{5}(2) = 2.7$
and it will be sorted behind $t_{i}^{10}(2)$ according to ACRC.
$L_{7} = L_{6} + l_{i}^{10} = 12.3 + 2.2 = 14.5$,
$L_{8} = L_{7} + l_{i}^{5} = 14.5 + 1.6 = 16.1$,
and $S_{6} = L_{6} * 2.9 = 35.67$, $S_{7} = L_{7} * 2.7 = 39.15$,
$S_{8} = L_{8} * 2.5 = 40.25$, then $S_{max} = S_{8}$.
If $R_{i}^{k}$ always equal to the most profitable revenue,
then $R_{i}^{2} = S_{8}$ and its unit price $p = 2.5$.
We assume that the matching result are the same in
stage \uppercase\expandafter{\romannumeral2},
then $m_{i}^{5}$ will be charged at the clearing
price $p_{i}^{5} = p_{i}^{5}(2) = l_{i}^{5} \cdot p = 1.6 * 2.5 = 4$ in stage \uppercase\expandafter{\romannumeral3}.
Then, if $m_{i}^{5}$ bids untruthfully, its utility is $\widetilde{u}_{i}^{5}$ where $\widetilde{u}_{i}^{5} = v_{i}^{5}(2) - p_{i}^{5} -
\theta = 4.96 - 4 - \theta = 0.96 - \theta$.
$m_{i}^{5}$ can improve its utility if the value of
the extra cost $\theta$ is small enough, e.g., $\theta < 0.96$,
when $m_{i}^{5}$ bids untruthfully.

\subsection{Scheme 3: TACDpp}

We introduce another efficient algorithm named TACDpp in this subsection.
The TACDpp is the improved version of TACDp, which refines
the second stage of TACDp.
The difference between TACDp and TACDpp is that, TACDpp
replaces algorithm FRM with algorithm FRMG in ASC.
The first stage of TACDpp is the same as that of  TACDp.
In the second stage, TACDpp matches cloudlets for APs in
a global way, which is different with TACDp.
In TACDpp, we match cloudlets with APs by algorithm
FRMG, which is shown in Algorithm \ref{alg:FRMG}.
Let $top_2$ be a small number, it is the top factor
in FRMG, its default value is $2$.
For each round, FRMG gets a random integer $rnd$ in $[1, top_2]$,
then selects the $rnd$th profitable value $d_{i}^{k}$
from the profit matrix $D$, and it returns $\{a_{i}, C_{k}\}$
to ASC for further judgement.
When the network is unbalanced between supply and
demand, i.e., $K \neq n$, TACDpp can perform
better due to the global idea.
It is also worth to mention that we must ensure
$top_2 > 1$ which is similar with $top_1$.
It will be discussed later.

\added{The performance comparison of the proposed
schemes is shown in Table \ref{table:Comparison}.
This table lists the algorithms employed in each stage
and the generation approach of the number $m$.}

\vskip -3mm
\begin{algorithm}[h]
    \small
	\begin{algorithmic}[1]
		\caption{FRMG: Finding a Rational Matching in the Global scope}\label{alg:FRMG}
    \REQUIRE{$D$, $top_2$}
    \ENSURE{$a_{i}$, $C_{k}$}
    \IF{$top_2 > 1$}
        \STATE{$rnd$ is the random integer in $[1, top_2]$}
    \ELSE
        \STATE{$rnd = 1$.}
    \ENDIF
    \STATE{Finding out the $rnd$-th profitable matching $d_{i}^{k}$ from $D$.}
    \RETURN{$a_{i}$, $C_{k}$}
    \end{algorithmic}	
\end{algorithm}
\vskip -3mm

\begin{table}[htbp]\scriptsize
\vskip -3mm
 \centering
 \caption{\label{table:Comparison}Comparison for TACD, TACDp and TACDpp}
 \begin{tabular}{C{1.2cm}C{1.5cm}C{2.5cm}C{1.5cm}}
  \hline\noalign{\smallskip}
  \textbf{Schemes}&Stage \uppercase\expandafter{\romannumeral1} & The number $m$ & Stage \uppercase\expandafter{\romannumeral2}  \\
  \noalign{\smallskip}\hline\noalign{\smallskip}

\textbf{TACD}&\textbf{ACRC}+\textbf{GTR}&[$(s+1)/2, s-1$]&\textbf{ASC}+\textbf{FRM}\\
\textbf{TACDp}&\textbf{ACRC}+\textbf{GTR}&One of $top_1$ cases&\textbf{ASC}+\textbf{FRM}\\
\textbf{TACDpp}&\textbf{ACRC}+\textbf{GTR}&One of $top_1$ cases&\textbf{ASC}+\textbf{FRMG}\\

 \noalign{\smallskip}\hline

 \end{tabular}
\vskip -3mm
\end{table}

\section{Desired Properties}
\label{sec:4}

\subsection{Truthfulness}

\begin{Theo}
The schemes TACD, TACDp and TACDpp are truthful in ACRC.
\begin{proof}
To verify the truthfulness of ACRC, we only need to prove
that MUs are truthful in our auction.
In TACD, for the MU $m_{i}^{j}$, $b_{i}^{j}(k)$ is the
truthful bid of $m_{i}^{j}$. Let $\widetilde{b}_{i}^{j}(k)$
be the untruthful bid.
Then the utility of $m_{i}^{j}$ is $u_{i}^{j}$ when
it bids truthfully. Let $\widetilde{u}_{i}^{j}$ be
the utility when it bids untruthfully.
We prove that $m_{i}^{j}$ cannot  improve its utility
by submitting an untruthful bid as follows, i.e., $\widetilde{u}_{i}^{j} \leq u_{i}^{j}$.

There are four cases for MU $m_{i}^{j}$ in TACD:
\begin{enumerate}
\item MU $m_{i}^{j}$ fails in the auction both in truthful
      bid $b_{i}^{j}(k)$ and untruthful bid $\widetilde{b}_{i}^{j}(k)$.
      Then, $u_{i}^{j} = 0$ and $\widetilde{u}_{i}^{j} = -\theta$.

\item MU $m_{i}^{j}$ wins the auction while bid truthfully
      and fails in the auction while bid untruthfully.
      In this case, $u_{i}^{j} \geq 0$, and $\widetilde{u}_{i}^{j} = -\theta$.

\item MU wins the auction both in truthful bid and untruthful bid.
      When $m_{i}^{j}$ wins the auction in TACD, its clearing price is $c$
      in our rules. On the other hand, if $m_{i}^{j}$ also wins the auction in
      another bid, from the definition of truthfulness,
      the clearing price is also $c$ while other  bids of MUs are fixed.
      Then $\widetilde{u}_{i}^{j} = u_{i}^{j} - \theta$.
\item MU fails in the auction while bid truthfully and wins
      the auction while bid untruthfully. When $m_{i}^{j}$ fails
      in TACD and it bids truthfully, the clearing price $c$ is
      greater than or equal to its bid, i.e., $c \geq b_{i}^{j}(k)$.
      And if $m_{i}^{j}$ wins the auction in another bid
      $\widetilde{b}_{i}^{j}(k)$, it must have $\widetilde{b}_{i}^{j}(k) \geq c$,
      so $\widetilde{b}_{i}^{j}(k) > b_{i}^{j}(k)$ and $\widetilde{b}_{i}^{j}(k) > v_{i}^{j}(k)$, then we have $\widetilde{u}_{i}^{j} \leq u_{i}^{j} = 0$.
\end{enumerate}

We have now discussed the truthfulness of MUs in TACD,
while MU $m_{i}^{j}$ bid for the $k$th cloudlet.
And the other cloudlets do not need care about
whether $m_{i}^{j}$ cheat or not, if the $k$th
cloudlet $C_{k}$ is assigned to the AP $a_{i}$ finally.

Similarly, MUs in TACDp and TACDpp are also truthful in ACRC,
because these two schemes only change the way we get the random integer $m$.
\end{proof}
\end{Theo}

\begin{Theo}
The schemes TACD, TACDp and TACDpp are truthful in ASC.
\begin{proof}

For TACD and TACDp, their algorithms in the second stage are similar to
the algorithm \textit{fixed price auction} as mentioned
in\cite{Goldberg2010Competitive}.
This auction scheme has been proved to be truthful, we only
change the generation manner of clearing price in TACD
and TACDp while the transactions is done.
Furthermore, the clearing price  is independent to AP and cloudlet
in the second stage of TACD and TACDp.
Therefore, TACD and TACDp are also truthful for ASC.

For TACDpp in ASC, we ensure its truthfulness by the top
factor $top_2$, which is discussed in the simulation section.
\end{proof}
\end{Theo}

\subsection{Budget Balanced}

\begin{Theo}
The schemes TACD, TACDp and TACDpp are budget balanced.
\begin{proof}

In this paper, we only prove that TACD is budget balanced.
The proof of TACDp and TACDpp are identical to that of  TACD.

In TACD, if $\sigma(i) = k$, $a_{i} \in W$ and
$C_{k} \in W'$, then the total clearing price charge
for the MUs is $val_{1}$ where $val_{1} = \sum_{i=1}^{n}\sum_{j=1}^{n_{i}}p_{i}^{j}$.
Similarly, the total clearing price for cloudlets
is $val_{2}$ where $val_{2} = \sum_{k=1}^{K}P^{k}$,
the total clearing price for APs is $val_{3}$
where $val_{3} = \sum_{i=1}^{n}(R_{i}^{k} - P_{i})$.
The total budget of APs is $val_{4}$ where
$val_{4} = \sum_{i=1}^{n}B_{i}^{k}$, then $val_{1} = val_{4}$
according to ACRC, and $val_{4} = val_{2} + val_{3}$ according to ASC.
Then, $val_{1} = val_{4} = val_{3} + val_{2}$, and
$val_{1} \geq val_{3} + val_{2}$,  i.e.,
 $$\sum_{i=1}^{n}\sum_{j=1}^{n_{i}}p_{i}^{j} \geq \sum_{k=1}^{K}P^{k} + (\sum_{i=1}^{n}R_{i}^{k} - \sum_{i=1}^{n}P_{i}).$$
\end{proof}
\end{Theo}

\subsection{Individual Rationality}
\begin{Theo}
The schemes TACD, TACDp and TACDpp are subject to the individual rationality.
\begin{proof}

The individual rationality for TACD can be proved as follows.
For sellers, according to the judgement in ASC, the
clearing price for cloudlets cannot smaller than
they asked, i.e., $P^{k}$ is always bigger than $r_{k}$.

For buyers, if MU $m_{i}^{j}$ wins the cloudlet
$C_{k}$, the MU will be charged at $p_{i}^{j}$
where $p_{i}^{j}= p \cdot l_{i}^{j}$.
$p$ is the performance price ratio of the $m$th MU
in $A$, and $p \leq t_{i}^{j}(k)$. Therefore,
$p_{i}^{j} \leq t_{i}^{j}(k) \cdot l_{i}^{j} = b_{i}^{j}(k)$.

For APs, we obtain $B_{i}^{k} = R_{i}^{k}$ according to the ACRC.
Also, the adjustment factor $f$ is in the scope of
$(0, 1)$ in ASC, thus, the clearing price of AP $P_{i} = f \cdot B_{i}^{k} < B_{i}^{k} = R_{i}^{k}$.
Therefore, $R_{i}^{k} \geq B_{i}^{k} \geq P_{i}$.

The proof of individual rationality for TACDp
and TACDpp iss the same as that of TACD.
\end{proof}
\end{Theo}

\subsection{Computational Efficiency}
\begin{Theo}
The time complexity of TACD as well as TACDp is $O(K\cdot n\log n)$.
\begin{proof}

For ACRC, the sorting needs $O(n\log n)$ time,
finding the number $s$ takes $O(n)$ time,
and the algorithm GTR also takes $O(n)$ time.
The time complexity of ACRC in TACD and TACDp
is $O(K\cdot n\log n)$.
For ASC, distributing APs randomly  takes $O(n\log n)$
time, the algorithm FRM takes $O(K)$ time. So,
the time complexity of ASC in TACD and TACDp is $O(n\cdot K)$.
Therefore the total time complexity of TACD and TACDp are $O(K\cdot n\log n)$.

\end{proof}
\end{Theo}

\begin{Theo}
The time complexity of TACDpp is $O(K\cdot n^{2})$.
\begin{proof}

The time complexity of ACRC in TACDpp is the
same as that of  TACDp, which is $O(K\cdot n\log n)$.
For ASC, the algorithm FRMG takes $O(n\cdot K)$ time,
which is different from the algorithm FRM. Thus,
the time complexity of ASC in TACDpp is $O(K\cdot n^{2})$.
Therefore the total time complexity of TACDpp is $O(K\cdot n^{2})$.
\end{proof}
\end{Theo}

\section{Numerical Results}
\label{sec:5}
\subsection{Simulation Setup}
In this paper, we simulate our works on MATLAB R2014a.
In the simulation, the capacities of all the cloudlets follow the normal distribution $N(25, 5)$
and each capacity $Cap^k$ satisfy the constraint $10 \leq Cap^{k} \leq 30$.
Its cost factor $c(k)$ follows to the normal
distribution $N(0.75, 0.1)$ and $0.5 \leq c(k) \leq 1$.
Then, its reserve price $\{r_{k}\}$ can be calculated
by formula \ref{fomul:rk}.
For each AP such as $a_{i}$, the number of MUs in $a_{i}$
follows the uniform distribution $U(5, 30)$.
For the MUs in $a_{i}$ such as $m_{i}^{j}$, their workload
follow the normal distribution $N(2, 1)$ and $1 \leq l_{i}^{j} \leq 3$.
Their valuations for each cloudlet follow
the uniform distribution $U(1, 15)$.

We compare our auction schemes with the strategy
Heaviest Access Point First (HAF) \cite{Jia2015Optimal}.
HAF is an efficient scheme for cloudlet placement
and resource allocation without auction.
In this paper, the strategy HAF is working in the following way, at first, HAF sorts APs in terms of the
total workload of MUs in descending order.
Then, HAF sorts cloudlets in terms of their capacity
in descending order.
At last, HAF matches cloudlets for APs by turns.
For instance, HAF assigns the first cloudlet whose
capacity is the biggest to the first AP whose total
workload of MUs is the heaviest, then HAF assigns
the second cloudlet to the second AP and so on.
If $C_{k}$ is assigned to $a_{i}$, the budget that
$a_{i}$ bid for $C_{k}$ is $B_{i}^{k}$.
It is calculated using the method as in ACRC, but the number $m$
is a fixed integer where $m = s$, and the potential
winner MUs is the first $m$ MUs in $A$, i.e., $A_{m}$.
The unit price $p$ charged by AP is the performance
price ratio of the $m$th MU in $A$.
In HAF, $a_{i}$ only needs to  calculate the budget on $C_{k}$.
The transaction between $a_{i}$ and $C_{k}$ will be done
if $B_{i}^{k} \geq r_{k}$, which is different from the algorithm ASC.
It is obvious that, if HAF is an incentive mechanism,
then it is untruthful.
\added{
Moreover, the time complexity
of HAF is $O(n\log n) + O(K\log K)$. In the first
stage of HAF, the sorting of APs and cloudlets takes
$O(n\log n)$ and $O(K\log K)$ time, respectively.
In the second stage of HAF, the matching algorithm
takes $O(n) + O(K)$ time.
       }

\subsection{Simulation Results}

\begin{figure}[htbp]
\vskip -3mm
\centering
\begin{minipage}{2in}
\includegraphics[width=2in]{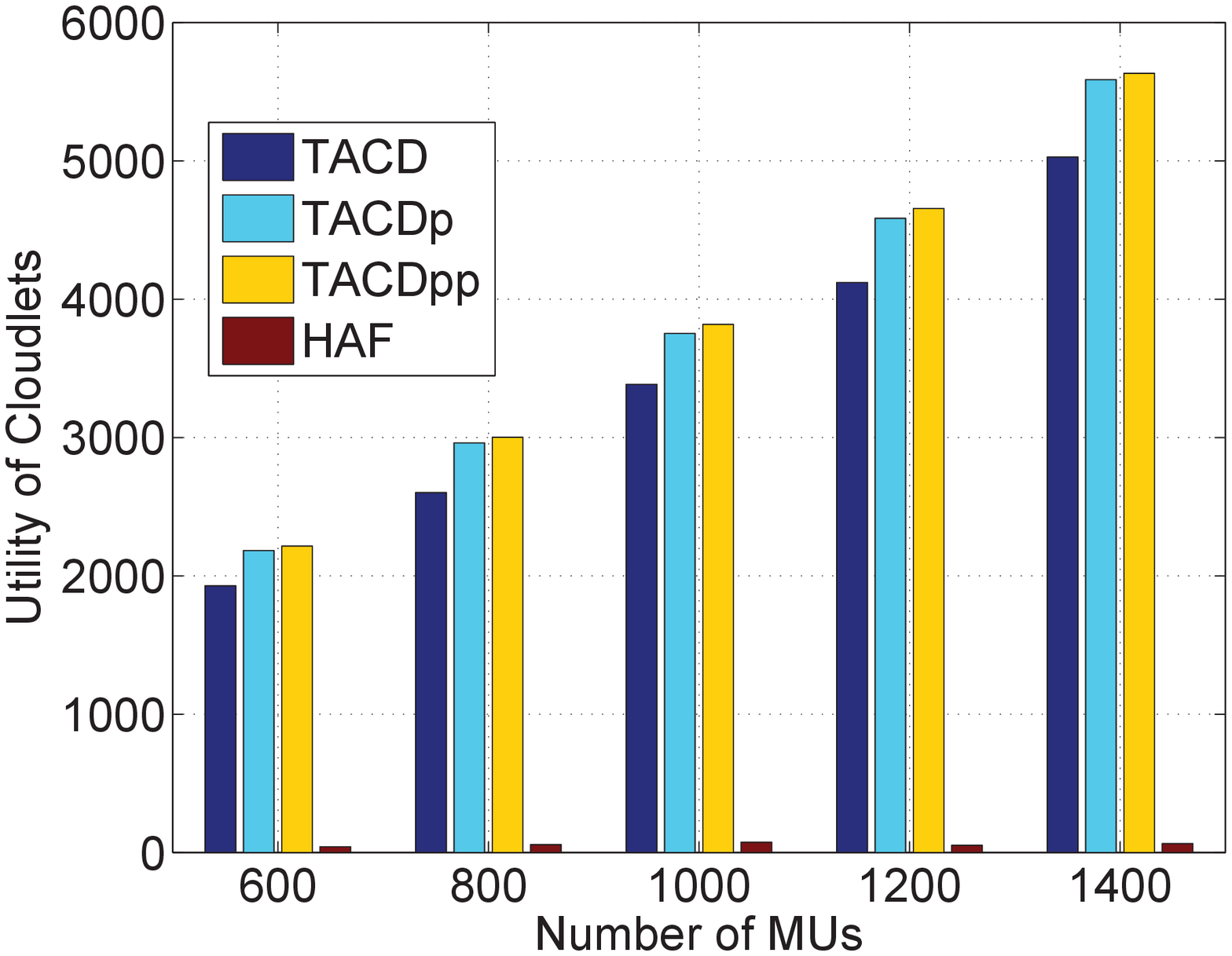}
\caption{\label{fig:utility_Cloudlet} {\small   Utility of Cloudlet.}}
\end{minipage}
\begin{minipage}{2in}
\includegraphics[width=2in]{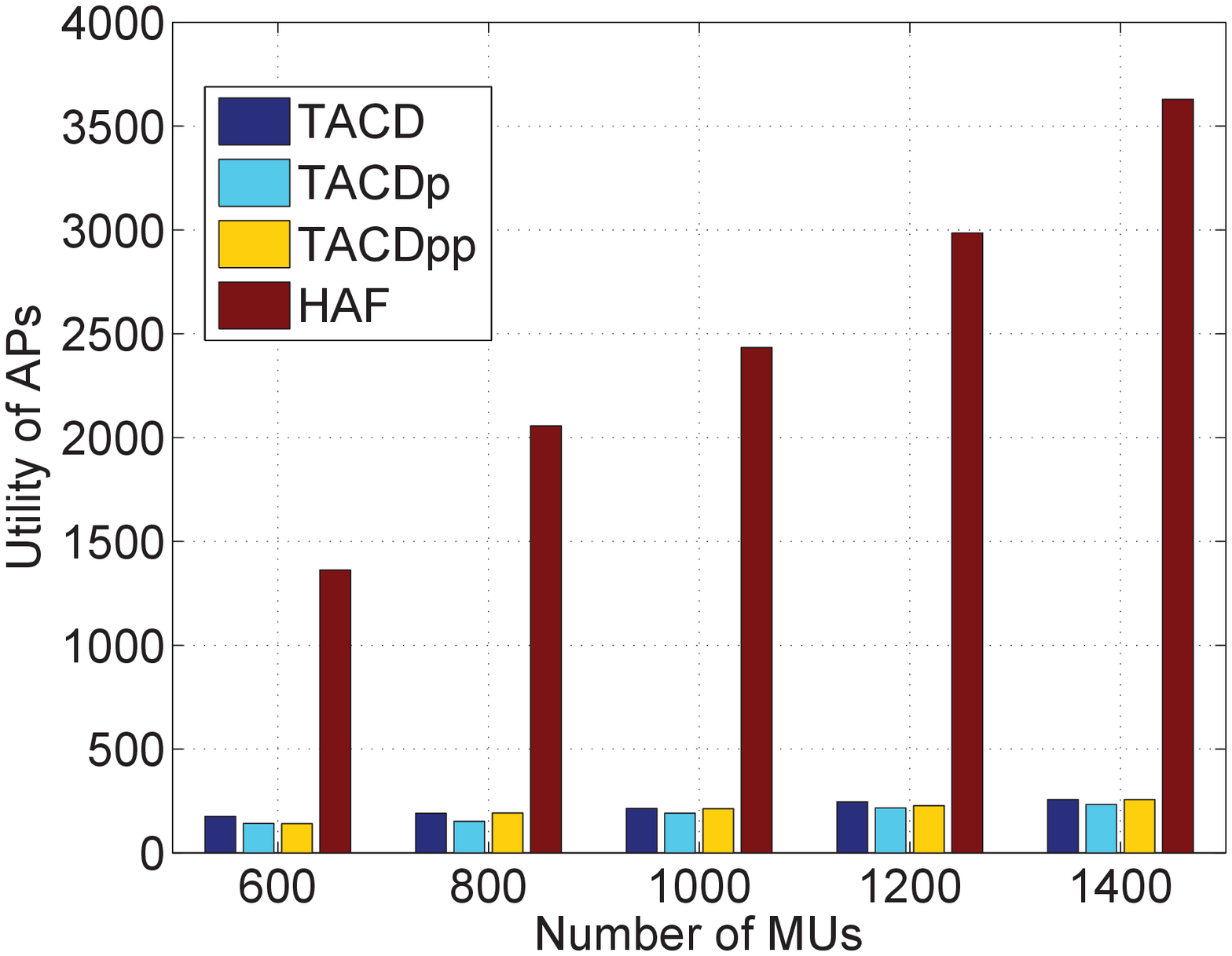}
\caption{\label{fig:utility_AP} {\small   Utility of APs.}}
\end{minipage}
\vskip -3mm
\end{figure}

\begin{figure}[htbp]
\vskip -1mm
\centering
\begin{minipage}{2in}
\includegraphics[width=2in]{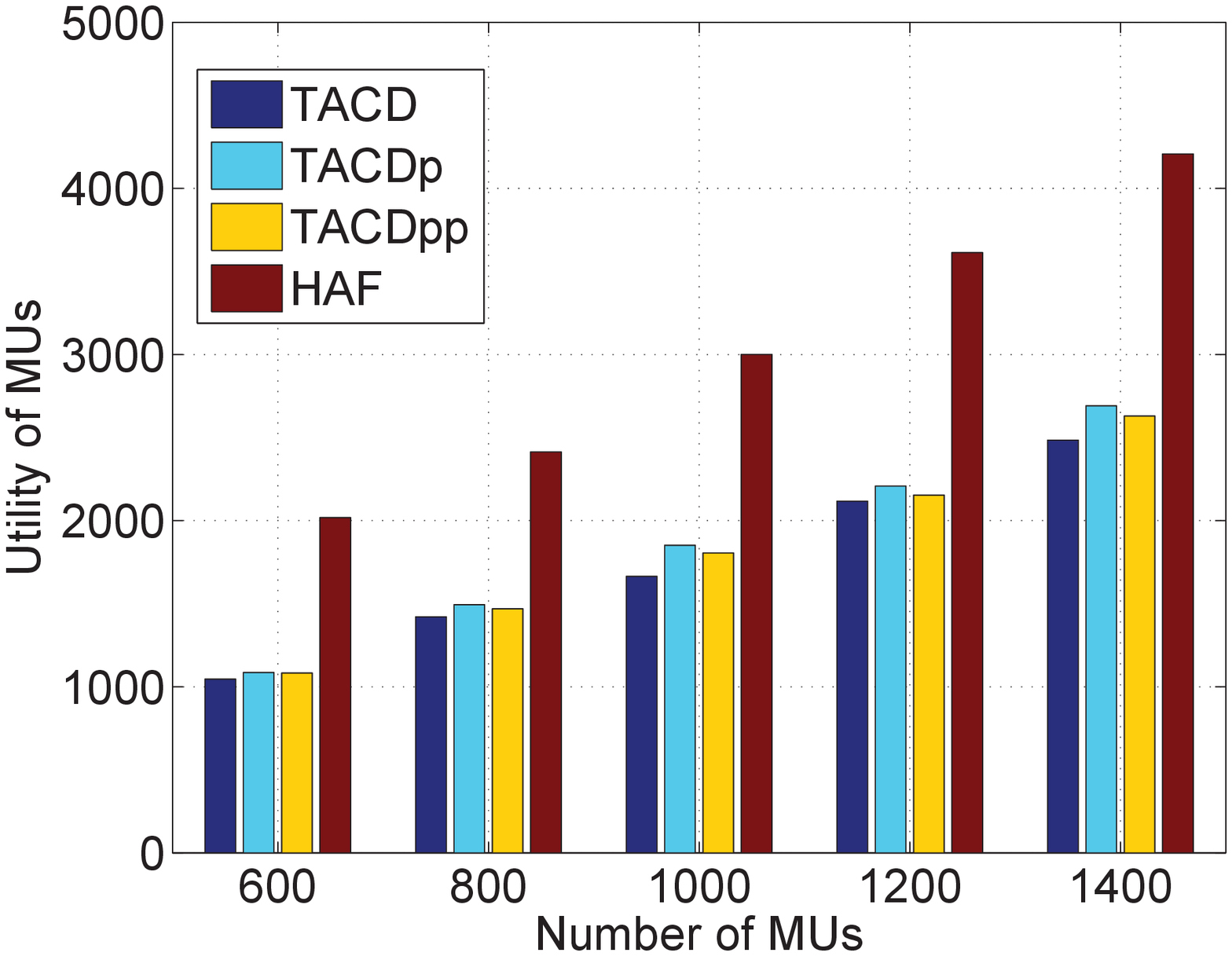}
\caption{\label{fig:utility_MU} {\small   Utility of MUs.}}
\end{minipage}
\begin{minipage}{2in}
\includegraphics[width=2in]{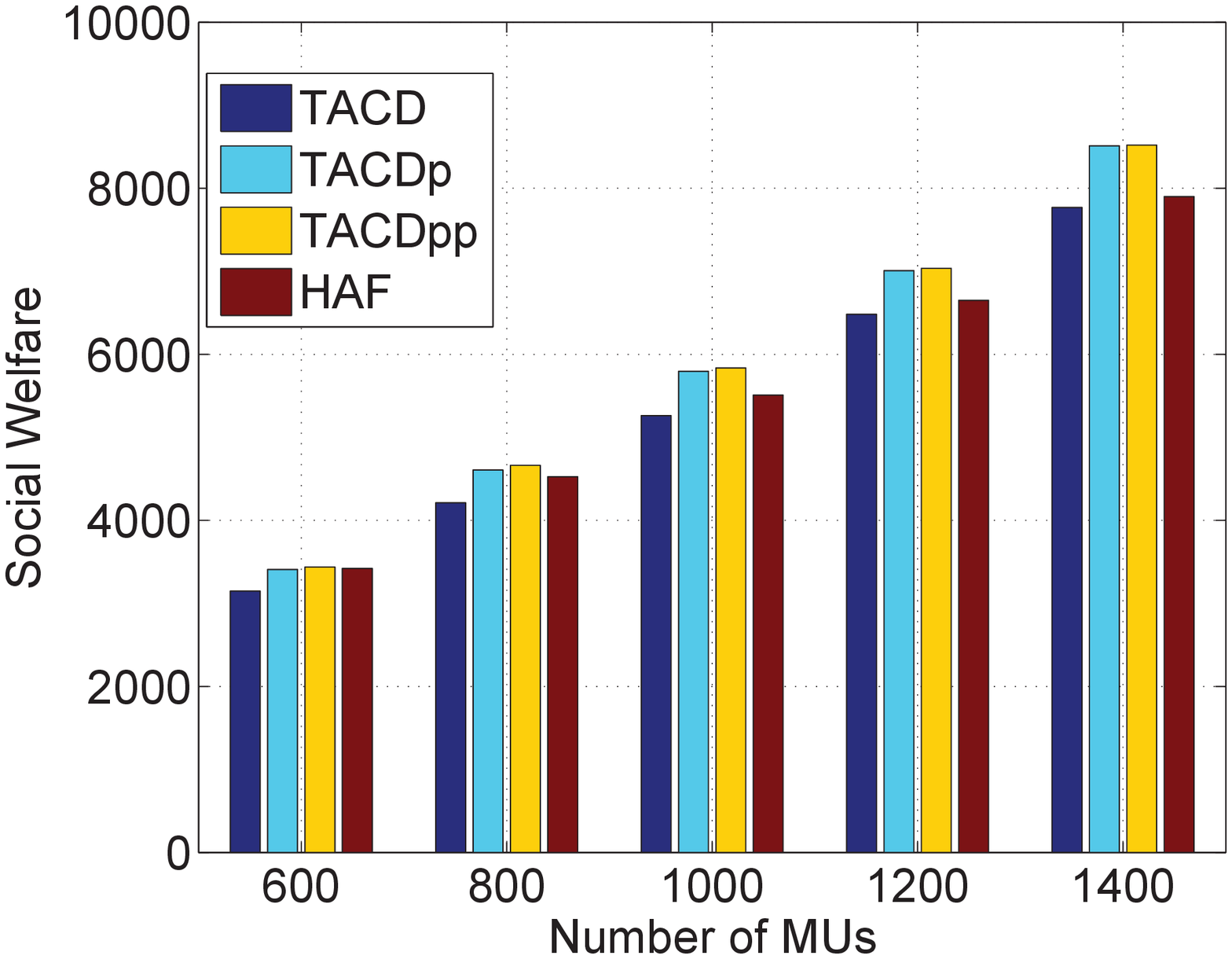}
\caption{\label{fig:Social_welfare} {\small   Social welfare.}}
\end{minipage}
\vskip -3mm
\end{figure}

In the first part of our simulation, the top factors
$top_1$ and $top_2$ are set to $2$, and the market
is balanced, i.e., $K = n$.
The utility of cloudlets, APs and MUs are shown in
Fig. \ref{fig:utility_Cloudlet}, Fig. \ref{fig:utility_AP},
and Fig. \ref{fig:utility_MU}, respectively.
The social welfare of auction schemes are shown in
Fig. \ref{fig:Social_welfare}.
There are big differences between our schemes and
the HAF in the first three figures.
Fig. \ref{fig:utility_Cloudlet} shows that our schemes
are good for cloudlets, while Fig. \ref{fig:utility_AP} show
that our schemes are weak for APs.
The differences are caused due to the following reasons.
In our schemes, we select a bid $B_{j}^{k}$ other
than $B_{i}^{k}$ and $r_{k}$ to keep ASC truthful where $B_{i}^{k} \geq B_{j}^{k} \geq r_{k}$.
The clearing price of this transaction
is $B_{j}^{k}$ which is bigger than $r_{k}$.
However, if $C_{k}$ is assigned for $a_{i}$,
HAF does not care about the truthfulness,
the transaction is done while $B_{i}^{k} \geq r_{k}$,
and the clearing price is equal to $r_{k}$.
As a result, the utility of cloudlets is close
to $0$ in HAF as shown in Fig. \ref{fig:utility_Cloudlet}.
Moreover, the APs in HAF may catch many profits
during the transaction as shown in Fig. \ref{fig:utility_AP}.
For the Fig. \ref{fig:utility_MU}, it shows that HAF is
more profitable for MU than our algorithms.
It is because that the winner cloudlet in HAF serve
for more MUs by a greedy manner and these MUs are
charged with a lower unit price by AP than our schemes.
In our schemes, the number of winner MUs is $m - 1$
where $m \leq s$, the unit price of these MUs is
the performance price ratio of the $m$th MU in $A$.
However, the number of winner MUs in HAF is $m$
where $m = s$, and the unit price of these MUs are
the performance price ratio of the $s$th MU in $A$.
Then, the number of winner MUs in HAF is more than
our schemes, and these winner MUs are charged by
a lower price than us.
Therefore, it is more profitable for MUs as show in Fig. \ref{fig:utility_AP}.
\added{The social welfare demonstrates that,
while the number of MUs is $1000$, the social welfare
in TACD is $5\%$ less than HAF, TACDp is $4.5\%$ higher
than HAF, and TACDpp is $5.6\%$ higher than HAF.
Moreover, our schemes perform better if there are more
MUs in the wireless access network.
For example, when the number of MUs is $1400$,
TACD is $1.7\%$ less than HAF, TACDp and TACDpp
are $7.6\%$ and $7.9\%$ higher than HAF respectively.}

\begin{figure}[htbp]
\vskip -3mm
\centering
\begin{minipage}{2in}
\includegraphics[width=2in]{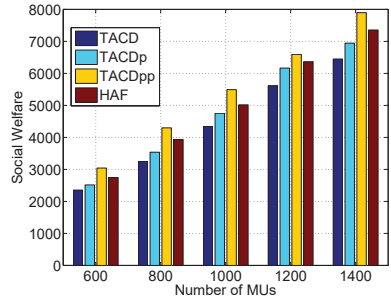}
\caption{\label{fig:Social_welfare_unbalanced} {\small Performance in unbalanced market.}}
\end{minipage}
\vskip -3mm
\end{figure}

If the number of APs is bigger than the number
of cloudlets, i.e., $n > K$, the performance of
our auction schemes in ``unbalanced market" is shown in Fig. \ref{fig:Social_welfare_unbalanced}.
In this situation, the TACDpp performs better than
that in the balanced market, because the global matching algorithm FRMG works better.

\begin{figure}[htbp]
\vskip -3mm
\centering
\begin{minipage}{2in}
\includegraphics[width=2in]{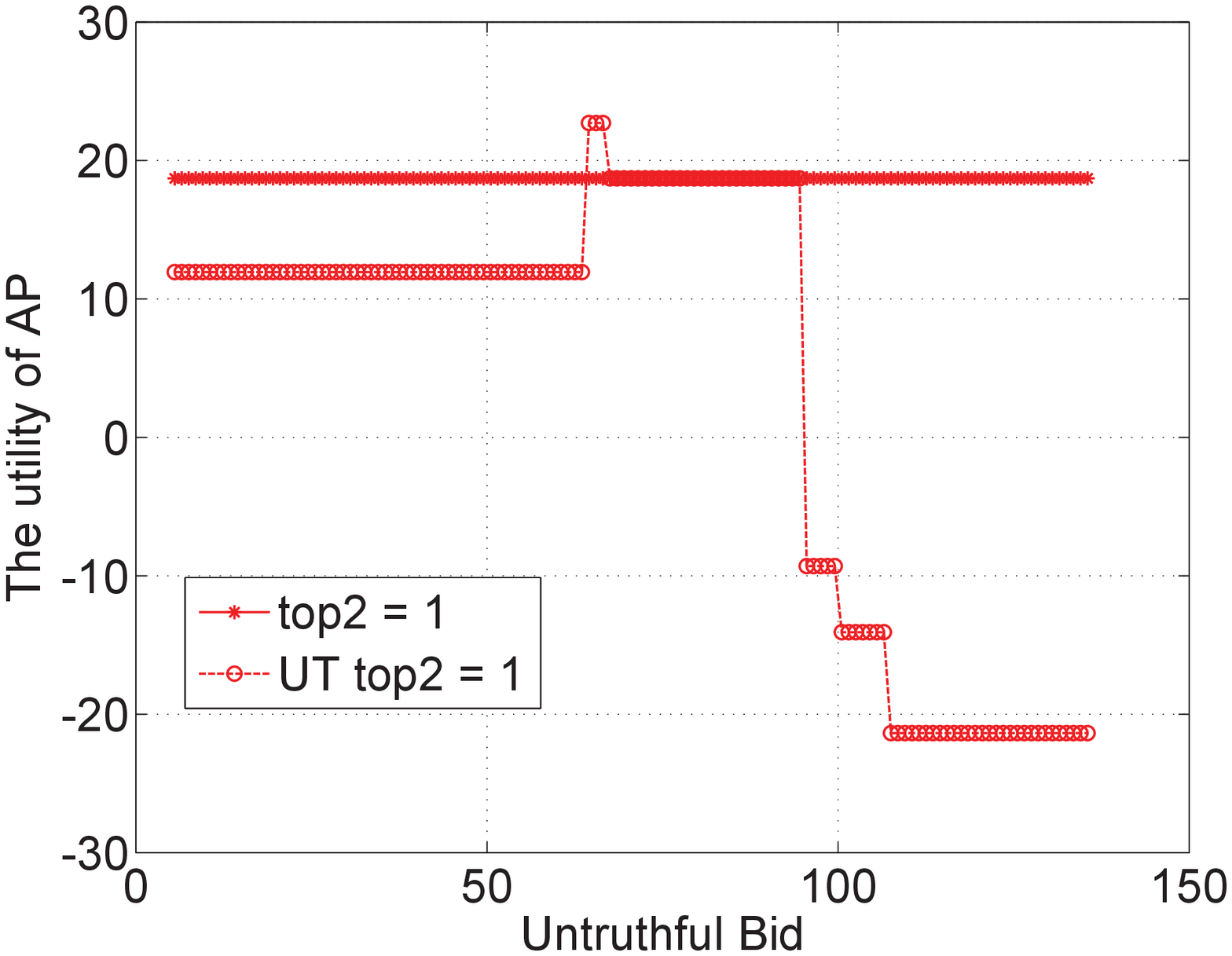}
\caption{\label{fig:top2=1} {\small  $top_2 = 1$.}}
\end{minipage}
\begin{minipage}{2in}
\includegraphics[width=2in]{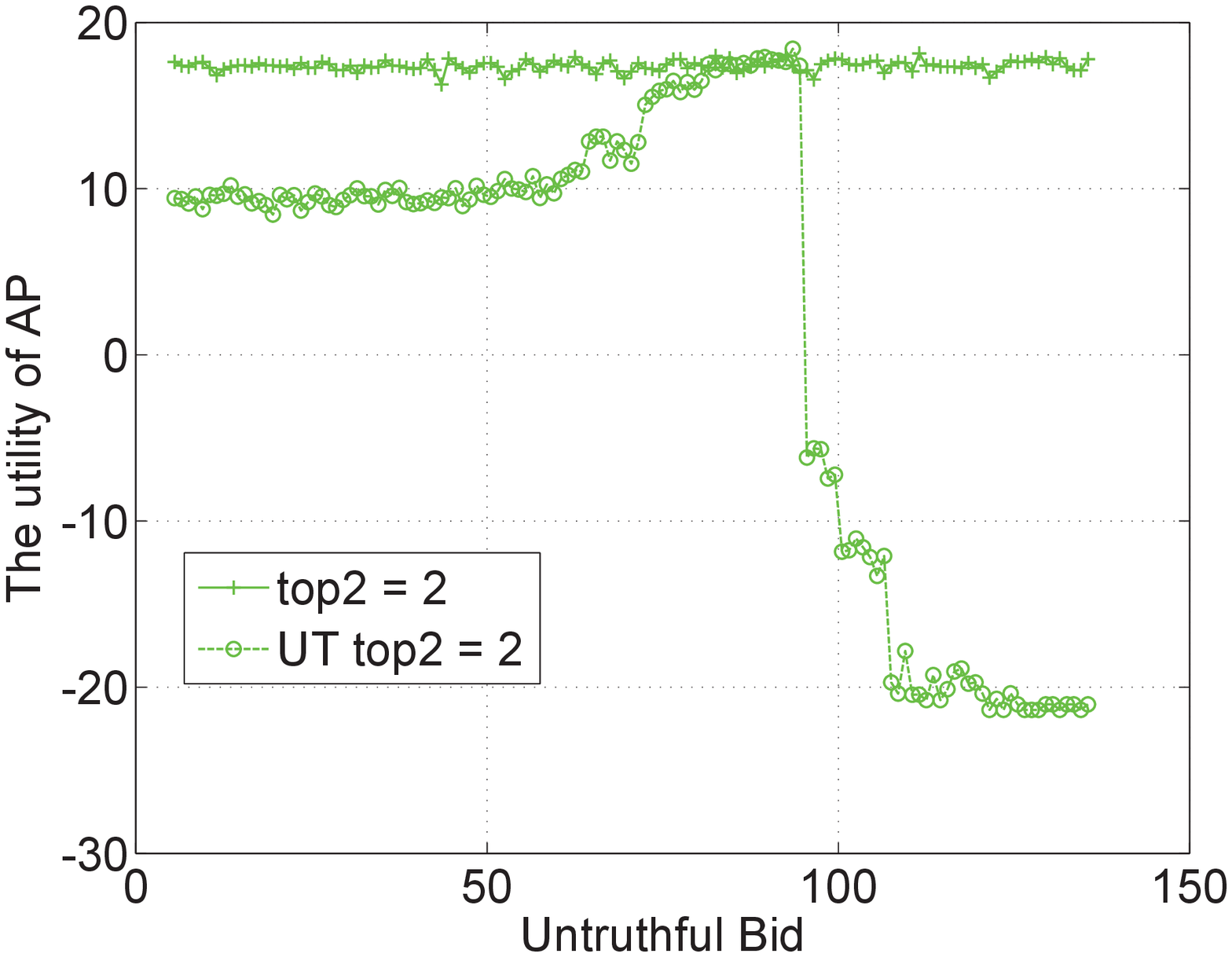}
\caption{\label{fig:top2=2} {\small $top_2 = 2$.}}
\end{minipage}
\vskip -3mm
\end{figure}

\begin{figure}[htbp]
\vskip -1mm
\centering
\begin{minipage}{2in}
\includegraphics[width=2in]{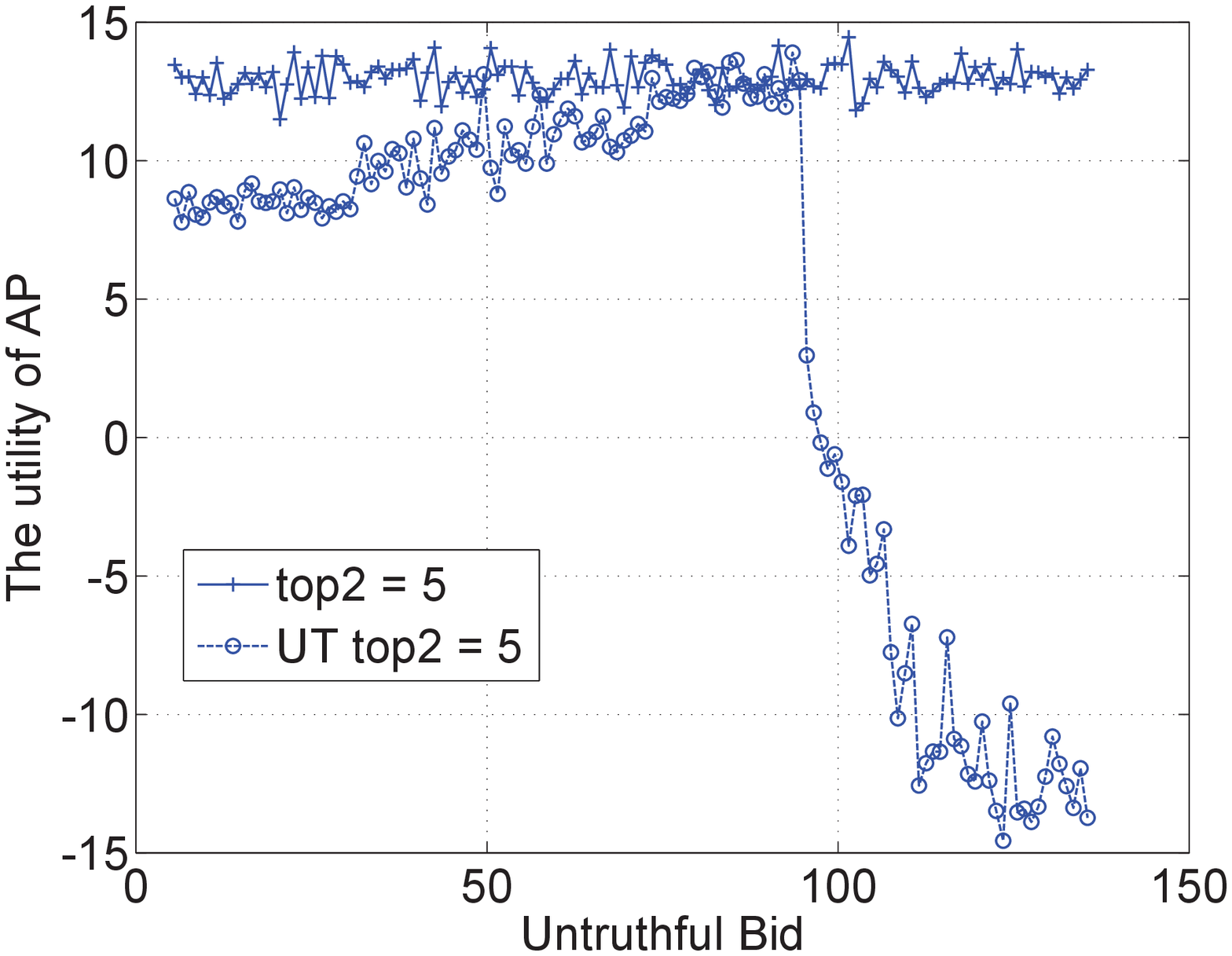}
\caption{\label{fig:top2=5} {\small  $top_2 = 5$.}}
\end{minipage}
\begin{minipage}{2in}
\includegraphics[width=2in]{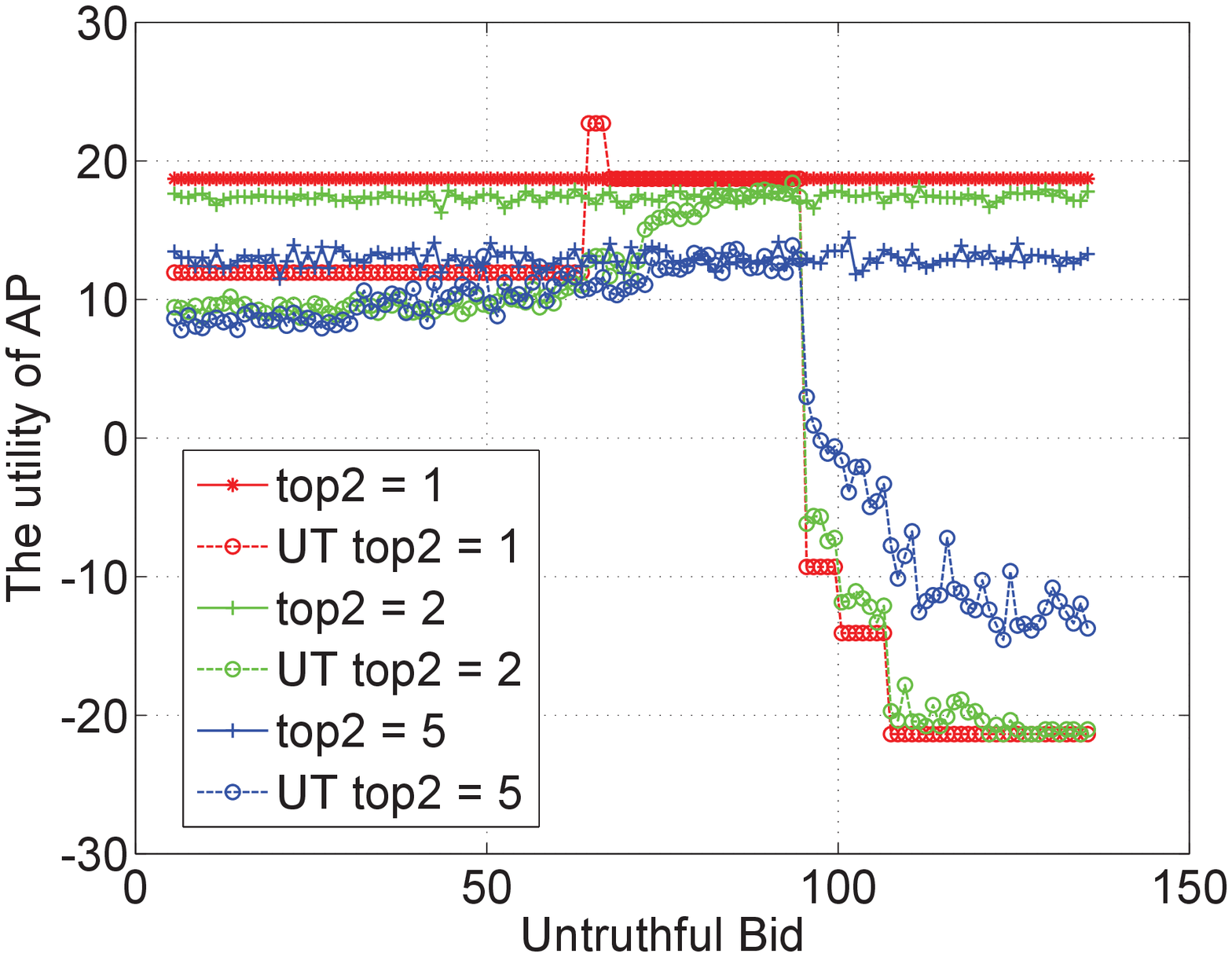}
\caption{\label{fig:top2=125} {\small Comparison.}}
\end{minipage}
\vskip -3mm
\end{figure}

Now, we evaluate the second stage of TACDpp while modifying
the value of $top_2$ in a smaller data set,
and we verify the truthfulness of TACDpp through
different values of $B_{1}^{1}$.
In this section, we fix the value of $top_1$ at $2$ and
modify the value of $top_2$ from $1$ to $2$ and then to $5$.
The utility of $B_{1}^{1}$  are shown in
Fig. \ref{fig:top2=1}, Fig. \ref{fig:top2=2}
and Fig. \ref{fig:top2=5}, for the cases of
$top_2 = 1, 2, 5$, respectively.
In these figures, the solid line shows the profit of
$a_{1}$ when $a_{1}$ bids truthfully in ASC,
i.e. $B_{1}^{1} = 85.5$.
The dotted line shows the
profit of $a_{1}$ when it bids untruthfully
from $\widetilde{B}_{1}^{1} = B_{1}^{1} - 80$
to $\widetilde{B}_{1}^{1} = B_{1}^{1} + 50$
with the increase unit of $1$.
The result is the averaged over $100$ random instances.
Fig. \ref{fig:top2=1} is the utility of AP for the case of
$top_2 = 1$. In this case, TACDpp matches cloudlet $C_{k}$
for AP $a_{i}$, while the profit of this matching
is the most profitable one in the rest of cloudlets and APs.
The utility of $a_{1}$ is $U_{1}$ and it is $18.7$.
It is stable and profitable, because TACDpp always makes
the same strategy to match cloudlets with APs.
In such a fixed strategy, $a_{1}$ will get the same profit
if it bids truthfully, so the solid line is straight in Fig. \ref{fig:top2=1}.
However, it is hard to check whether TACDpp is truthful in ASC while $top_2 = 1$.
Because it may has some "bugs", in which APs can
benefit more from their preferred cloudlet,
by biding budgets that lower than their revenues.
For instance, as we can see in Fig. \ref{fig:top2=1},
the  utility of $a_{1}$  is $\widetilde{U}_{1}$ where
$\widetilde{U}_{1} = 22.7 - \theta$, while $a_{1}$ bid untruthfully
among $\{64.5, 65.5, 66.5\}$.
$\widetilde{U}_{1}$ is larger than $U_{1}$, if $\theta < 4$.
It is because that, when $\widetilde{B}_{1}^{1}
= \{64.5, 65.5, 66.5\}$, the profit $\widetilde{B}_{1}^{1} - r_{1}$
is so big  that $a_{1}$ still wins $C_{1}$.
Also, there is another AP $a_{x}$ whose budget is
$B_{x}^{1}$ where $B_{x}^{1} \leq 64.5$, and it is the
largest $B_{j}^{1}$ in which $B_{j}^{1} \leq
\widetilde{B}_{1}^{1}$, $j \in [1, n]$ and $j \neq 1$.
Then, the clearing price will be much lower than that when it bids truthfully.
Therefore, if AP $a_{1}$ pays  some extra price
$\theta$ to figure out these more profitable
cases, it will get more profits firmly by bidding untruthfully.

Simulation results of TACDpp are shown in
Fig. \ref{fig:top2=2} where $top_2 = 2$.
The solid line shows the utility of $a_{1}$
while it bids truthfully. This line is not
a straight line as shown in Fig. \ref{fig:top2=1},
as the matching strategy is
not a fixed pure strategy anymore.

When $top_2 = 2$, the matching strategy turns to
a mixed strategy, we combine the following  two strategies with
equal probability, i.e. $1/2$,

\begin{enumerate}
\item Matching cloudlet $C_{k}$ with AP $a_{i}$
      whose profit $B_{i}^{k} - r_{k}$ is the most profitable one.
\item Matching cloudlet $C_{k}$ with AP $a_{i}$
      whose profit $B_{i}^{k} - r_{k}$ is the second profitable one.
\end{enumerate}

So the utility of $a_{1}$ is not a stable value,
even $a_{1}$ always bid truthfully.
The utility varies within an interval near $18$,
which is shown in green solid line.
In contrast, the green dotted line shows the
utility of $a_{1}$ while it bids untruthfully.
There are also some more profitable cases while
$a_{1}$ bids untruthfully, such as $\{64.5, 65.5, 66.5\}$
as occour ed as in the case of $top_2 = 1$.
But the difference is that, if $top_2 = 2$,
$a_{1}$ can also benefit more in those cases
with the probability of $50\%$.
Otherwise, $a_{1}$ will be matched with other
less profitable cloudlets, and it also must
pay an extra cost $\theta$ to find those cases.
Therefore, there is not any evident case in
which $a_{1}$ can get more utility than the truthful case.
It is worthless for $a_{1}$ to pay an extra
cost $\theta$ to determine how to bid untruthfully.
Therefore, TACDpp is truthful while $top_2 = 2$.

Similarly, Fig. \ref{fig:top2=5} shows the utility of
$a_{1}$ while $top_2 = 5$.
It is also a mixed strategy by $5$ pure strategies,
with the probability of $1/5$ for each strategy.
These $5$ pure strategies are used to match
cloudlets to APs.
The $j$th pure strategy is corresponding to
the $j$th profitable value of $B_{i}^{k} - r_{k}$
for $j=1, 2, \cdots, 5$. In the mixed strategy,
the utility of $a_{1}$ varies with a larger
range than that in  Fig. \ref{fig:top2=2}
while $a_{1}$ bid truthfully.
The value of its utility fall in $[12, 14.5]$,
and it is less than that in Fig. \ref{fig:top2=2}.
In other words, the strategy for $top_2 = 5$ is
less profitable and less stable than the strategy
for $top_2 = 2$, while APs bid truthfully.
This is because the stronger randomness
brings APs many solutions which are not profitable.
For truthfulness, there is no evidence that $a_{1}$ can get more utility than the truthful one.

\section{Conclusion}
\label{sec:6}

In this paper, we have proposed efficient auction schemes for cloudlets placement and resource allocation in wireless networks to improve the social welfare subject to economic properties. We have introduced the group-buying model to inspire cloudlets to serve the MUs. In our auction schemes, MUs can get access to cloudlets through APs, according to their preference and resource demands for cloudlets. The whole three entities MUs, APs, and cloudlets are motivated to participate in resource sharing. We have verified that our schemes are truthful, individual rational, budget balanced and computational efficient. Through simulations, we have shown that our schemes TACDp and TACDpp outperform HAF by about $4.5\%$ and $5.6\%$ respectively, in terms of social welfare, for the case that the number of MUs is 1000.

% BibTeX users please use one of
%\bibliographystyle{spbasic}      % basic style, author-year citations
%\bibliographystyle{spmpsci}      % mathematics and physical sciences
%\bibliographystyle{spphys}       % APS-like style for physics
%\bibliography{}   % name your BibTeX data base

% Non-BibTeX users please use

%\bibliographystyle{spmpsci}
\bibliographystyle{IEEEtran}
\bibliography{refs_r1}

\end{spacing}
\end{document}